\documentclass[twocolumn,showpacs,amsmath,amssymb]{revtex4}

\usepackage{epsf}
\usepackage{graphicx}
\usepackage{amsfonts}
\usepackage{dcolumn}%
\usepackage{color,ulem}
\usepackage{bm}

\begin{document}
 
\title{Magnetization induced by odd-frequency spin-triplet Cooper pairs in a Josephson junction with metallic trilayers} 

\author{S. Hikino$^{1}$}
\author{S. Yunoki$^{1,2,3}$}%
\affiliation{%
$^{1}$Computational Condensed Matter Physics Laboratory, RIKEN ASI, Wako, Saitama 351-0198, Japan \\
$^{2}$Computational Materials Science Research Team, RIKEN Advanced Institute for Computational Science (AICS), 
Kobe, Hyogo 650-0047, Japan\\
$^{3}$Computational Quantum Matter Research Team, RIKEN Center for Emergent Matter Science (CEMS), Wako, Saitama 351-0198, Japan
} 

\date{\today}

\begin{abstract}
We theoretically study the magnetization inside a normal metal induced in 
an $s$-wave superconductor/ferromagnetic metal/normal metal/ferromagnetic metal/$s$-wave superconductor ({\it S}/{\it F}1/{\it N}/{\it F}2/{\it S}) 
Josephson junction. 
Using the quasiclassical Green's function method, we show that the magnetization becomes finite inside the $N$. 
The origin of this magnetization is due to odd-frequency spin-triplet Cooper pairs 
formed by electrons of equal and opposite spins, 
which are induced by the proximity effect in the {\it S}/{\it F}1/{\it N}/{\it F}2/{\it S} junction. 
We find that the magnetization $M(d,\theta)$ in the {\it N} can be decomposed into two parts, 
$M(d,\theta)=M^{\rm I}(d)+M^{\rm II}(d,\theta)$, 
where $\theta$ is the superconducting phase difference between the two {\it S}s and $d$ is the thickness of $N$. 
The $\theta$ independent magnetization $M^{\rm I}(d)$ exists generally in $S$/$F$ junctions, while 
$M^{\rm II}(d,\theta)$ carries all $\theta$ dependence and represents the fingerprint of the phase coherence between the two {\it S}s 
in Josephson junctions. 
The $\theta$ dependence thus allows us to control the magnetization in the $N$ by tuning $\theta$ for  a fixed $d$. 
We show that the $\theta$ independent magnetization $M^{\rm I}(d)$ weakly decreases with increasing $d$, 
while the $\theta$ dependent magnetization $M^{\rm II}(d,\theta)$ rapidly decays with $d$. 
Moreover, we find that the time-averaged magnetization $\langle M^{\rm II}(d,\theta)\rangle$ exhibits a discontinuous peak at each resonance 
DC voltage $V_n = n\hbar\omega_{\rm S}/2e$ ($n$: integer) 
when DC voltage $V$ as well as AC voltage $v_{\rm ac}(t)$ with frequency $\omega_{\rm S}$ 
are both applied to the {\it S}/{\it F}1/{\it N}/{\it F}2/{\it S} junction. 
This is because $ M^{\rm II}(d,\theta)$ 
oscillates generally in time $t$ (AC magnetization) with $d\theta/dt = 2e[V+v_{\rm ac}(t)]/\hbar$ and thus $\langle M^{\rm II}(d,\theta)\rangle=0$, 
but can be converted into the time-independent 
DC magnetization for the DC voltage at $V_n$. 
We also discuss that the magnetization induced in the {\it N} can be measurably large in realistic systems. 
Therefore, the measurement of the induced magnetization serves as an alternative way to detect the phase coherence between 
the two {\it S}s in Josephson junctions. 
Our results also provide a basic concept for tunable magnetization in superconducting spintronics devices. 
\end{abstract}


\maketitle 


\section{Introduction}\label{sec:introduction}
The proximity effect is an important quantum phenomenon which occurs when a superconductor is attached to 
non-superconducting materials, 
where the pair amplitude of Cooper pairs in the superconductor penetrates into the non-superconducting materials~\cite{degennes}. 
A typical example is the Josephson effect, 
which has been known as one of the macroscopic quantum phenomena, 
characterized as DC current flowing without a voltage-drop between two superconductors separated by a thin insulator ({\it I}) 
or normal metal ({\it N})~\cite{josephson,likhalev}. 
The Josephson critical current in an superconductor/insulator/superconductor or superconductor/normal metal/superconductor junction 
forming a Josephson junction monotonically decreases with increasing the thickness of {\it I} or {\it N}~\cite{degennes,josephson,likhalev}.

The proximity effect in $s$-wave superconductor/ferromagnetic metal ({\it S}/{\it F}) hybrid junctions 
has been extensively studied in the last decade 
because of its fascinating phenomena and potential applications to superconducting spintronics~\cite{buzdin-jetp, ryazanov-prl, kontos-prl, sellier-pr, bauer-prl, frolov-prb, robinson, born-prb, weides-prl, oboznov-prl, shelukhin-prb, pfeiffer-prb, bannykh-prb, khaire-prb, wild-epjb, kemmer-prb, golubov-rmp, buzdin-rmp, bergeret-rmp,linder-prb90}. 
Due to the proximity effect between {\it S} and {\it F} in a {\it S}/{\it F} junction, the spin-singlet Cooper pairs (SSCs) penetrate into the {\it F} and acquire a finite 
center-of-mass momentum proportional to the exchange splitting between up- and down-spin bands in the {\it F}. 
The pair amplitude of SSC shows damped oscillation with increasing the thickness of {\it F}. 
One interesting phenomena induced by the damped oscillatory behavior of the pair amplitude of SSC is a $\pi$-state in a {\it S}/{\it F}/{\it S} junction, 
where the current-phase relation in the Josephson junction is shifted by $\pi$ from that of the ordinary {\it S}/{\it I}/{\it S} or {\it S}/{\it N}/{\it S} junction (called 0-state)
~\cite{buzdin-jetp, ryazanov-prl, kontos-prl, sellier-pr, bauer-prl, frolov-prb, robinson, 
born-prb, weides-prl, oboznov-prl, shelukhin-prb, pfeiffer-prb, bannykh-prb, khaire-prb, wild-epjb, kemmer-prb, golubov-rmp, buzdin-rmp, bergeret-rmp}. 
It is expected that the $\pi$-state can be utilized for an element of quantum computing and circuit~\cite{yamashita-qbit, bell-apl84, khabipov, hikino-jpsj84}. 

Another intriguing proximity effect in {\it S}/{\it F} hybrid junctions is the emergence of odd-frequency spin-triplet Cooper pairs (STCs), 
although the $S$ is an $s$-wave superconductor~\cite{bergeret-rmp, linder-np}. 
Here, the anomalous Green's functions of spin-triplet components are odd functions with respect to the fermion Matsubara 
frequency $\omega_{n}$. 
It should be noted that the anomalous Green's functions in bulk superconductors are generally even functions with respect to $\omega_{n}$. 
When the magnetization in the {\it F} is uniform in a {\it S}/{\it F} junction, 
not only the SSC, as described above, but also the STC composed of opposite spin electrons 
(i.e., total spin projection on $z$ axis being $S_z=0$) penetrates into the {\it F} due to the 
proximity effect~\cite{bergeret-rmp,yokoyama-prb75}. 
The penetration length of STC with $S_z=0$ (and also SSC) into the {\it F} is very short and 
the amplitude of STC exhibits a damped oscillatory behavior inside the {\it F} with increasing the thickness of {\it F}. 
The penetration length is determined by $\xi_{\rm F}=\sqrt{\hbar D_{\rm F}/h_{\rm ex}}$, which is typically a order of few nanometers~\cite{buzdin-jetp, ryazanov-prl, kontos-prl, sellier-pr, 
bauer-prl, frolov-prb, robinson, born-prb, weides-prl, oboznov-prl, shelukhin-prb, pfeiffer-prb, bannykh-prb, khaire-prb, 
wild-epjb, kemmer-prb, golubov-rmp, buzdin-rmp, bergeret-rmp}. 
Here, $D_{\rm F}$ and $h_{\rm ex}$ are the diffusion coefficient and the exchange field in the {\it F}, respectively. 

On the contrary, when the magnetization in the {\it F} is non-uniform in a {\it S}/{\it F} junction, 
the STC formed by electrons of equal spin ($|S_z|=1$) can also be induced in the {\it F}. 
This includes cases, for instance, where the $F$ contains a magnetic domain wall~\cite{bergeret-prl86, champel-prb72, braude-prl98, fominov-prb75, volkov-prb78, alidoust-prb81, bzdin-prb83}, the junction consists of {\it F} multilayers~\cite{volkov-prb90, bergeret-prb68, houzet-prb76, trifunovic-prb82, volkov-prb81, trifunovic-prb84, melnikov-prl109, 
pugach-apl101, knezevic-prb85, richard-prl110, fritsch-njp16, alidoust-prb89,fominov-jetpl77, Lofwander-prl95, halterman-prb77, fominov-jetpl91, kawabata-jpsj82, hikino-prl110, mironov-prb89}, the interface of $S$/$F$ junction is spin active~\cite{eschrig-sh, asano-prl, galaktionov-prb, beri-prb, linder-prb82, trifunovic-prl107}, 
and the ferromagnetic resonance occurs~\cite{takahashi-prl99, houzet-prl101, yokoyama-prb80}. 
Although the pair amplitude of STC with $|S_z|=1$ monotonically decreases with increasing the thickness of {\it F}, 
the STC with $|S_z|=1$ can propagate into the {\it F} over a distance of the order of 
$\xi_{\rm 0}=\sqrt{\hbar D_{\rm F}/2\pi k_{\rm B}T}$ ($T$: temperature), 
which is typically about several dozen nanometers~\cite{deutscher}. 
This is approximately 2 orders of magnitude longer than the penetration length of the SSC and the STC with $S_z=0$. 
Therefore, the proximity effect of STCs with $|S_z|=1$ is called the long-ranged proximity effect (LRPE).

Following the theoretical predictions, the STC in {\it S}/{\it F} hybrid junctions has been confirmed experimentally~\cite{keizer-n439, robinson-science, khaire-prl, anwar-apl, anwar, leksin-prl109, wang-prb89,bakurskiy-prb88}. 
The obvious way to observe the LRPE induced by the STC with $|S_z|=1$ is to directly measure the Josephson current in Josephson junctions 
composed of $F$s~\cite{keizer-n439, robinson-science, khaire-prl, anwar-apl, anwar}. 
Indeed, the LRPE has been observed in $S$/$F$ junctions with spin-active interfaces~\cite{keizer-n439, anwar-apl, anwar} 
and in $S$/$F$ multilayer systems with non-collinear magnetization alignment between $F$ layers~\cite{robinson-science, khaire-prl}. 
Recently, the variation of superconducting transition temperature ($T_{\rm C}$) has been observed in {\it S}/{\it F}1/{\it F}2 type spin valve structures 
as the direction of magnetizations in the two ferromagnetic metals $F$1 and $F$2 is changed~\cite{leksin-prl109,wang-prb89}. 
This is also due to the 
LRPE induced by the STC as predicted in the previous theoretical calculation~\cite{fominov-jetpl91}.

An alternative way to prove the STC is to measure the spin angular momentum carried by Cooper pairs 
because the spin is finite for the STC but is zero for the SSC. 
Several theoretical studies have already addressed this issue and examined the magnetization induced by the STC in 
the various geometry of {\it S}/{\it F} hybrid structures~\cite{bergeret-rmp,Lofwander-prl95,halterman-prb77,pugach-apl101}. 
A {\it F}/{\it S}/{\it F} junction with a spin valve structure is a typical geometry of such {\it S}/{\it F} hybrid structures. 
When the magnetizations in the two $F$s separated by the $S$ are non-collinearlly aligned, 
not only the STC with $|S_z|=0$ but also 
the STC with $|S_z|=1$ becomes finite and induces a finite magnetization 
inside the {\it S} as well as the two {\it F}s~\cite{bergeret-rmp,Lofwander-prl95,halterman-prb77}.

Recently, the magnetization induced by the STC has also been studied in Josephson junction type multilayer systems, e.g., 
{\it S}/{\it F}/{\it F}/{\it S}, {\it S}/{\it F}/{\it F}/{\it S}/{\it F}, and rather complex symmetric three terminal 
{\it S}/{\it F}/{\it F}/{\it S}/{\it F}/{\it F}/{\it S} junctions~\cite{pugach-apl101}. 
It has been pointed out that such Josephson junctions with metallic ferromagnetic multilayers, especially, the symmetric three terminal 
{\it S}/{\it F}/{\it F}/{\it S}/{\it F}/{\it F}/{\it S} junction may have promising potential for superconducting spintronics applications with low dissipation~\cite{pugach-apl101}. 
This is because the magnetization in this junction can be well controlled by changing the superconducting phase difference between the 
two outmost {\it S}s without Jule heating. 
Here, it should be noted that the thickness $d_{\rm S}$ of S in the middle layer sandwiched by the two ferromagnetic double layers 
has to be $d_{\rm S}\ll \xi_{\rm S}^{2}/\xi_{\rm 0}$ ($\xi_{\rm S}$: superconducting coherence length) 
in order to observe clearly the magnetization in the middle $S$ layer induced by the STC~\cite{pugach-apl101}. 
However, in this case, the superconductivity in the middle {\it S} layer is violently suppressed. To prevent this from happening, for example, a three 
terminal Josephson junction composed of large superconducting electrodes in the middle $S$ layer 
is proposed~\cite{pugach-apl101, alidoust-prb89, bakurskiy-prb88}.

In this paper, we focus on a much simpler Josephson junction with metallic trilayers, i.e., a {\it S}/{\it F}1/{\it N}/{\it F}2/{\it S} Josephson junction (see Fig.~\ref{sfnfs-gm}), which nowadays has been 
able to be fabricated experimentally~\cite{robinson-science, khaire-prl, iovan-prb90}, and theoretically examine, by employing 
the quasiclassical Green's function method, the magnetization 
inside the $N$ induced by the odd-frequency STCs composed of electrons of equal and opposite spins. 
Fixing the magnetization in $F2$ along the $z$ direction perpendicular to the junction direction 
($x$ direction), 
we show that i) the $x$ component of the magnetization in the $N$ is always zero, ii)   
the $y$ component becomes exactly zero when the magnetizations in $F$1 and $F$2 are 
collinear, and iii) the $z$ component is generally finite for any magnetization alignment between 
$F$1 and $F$2. 
We also show that the magnetization in the $N$ can be decomposed into two parts, $\theta$ dependent and independent parts, 
where $\theta$ is the superconducting phase difference between the two $S$s in the $S$/$F$1/$N$/$F$2/$S$ junction.  
The $\theta$ dependent magnetization is induced as a result of finite coupling between the two $S$s, 
while the $\theta$ independent magnetization always exists due to the proximity effect in $S$/$F$ hybrid junctions. 
We find that the $\theta$ independent magnetization decreases slowly with increasing the thickness of $N$, 
whereas the $\theta$ dependent magnetization decays rather rapidly. 
We also investigate the dynamics of the magnetization in the $N$ when AC voltage is applied. 
Because of the AC voltage, the superconducting phase difference $\theta$ is now time dependent and accordingly the $\theta$ dependent part of the magnetization oscillates. However,  
we find that the $\theta$ dependent part of the magnetization is converted, when it is time averaged, from the oscillating AC character to the time-independent 
DC character at specific DC voltages, depending on the frequency of AC voltage, if DC and AC voltages are both applied to 
the $S$/$F$1/$N$/$F$2/$S$ junction. Finally, we argue that the magnetization 
induced inside the $N$ can be large enough to be observed experimentally in realistic settings.

The rest of this paper is organized as follows. 
In Sec.~\ref{sec:formulation}, we introduce a simple $S$/$F$1/$N$/$F$2/$S$ junction consisting of  metallic trilayers and derive the analytical 
formulation of the magnetization induced inside the $N$ 
on the basis of Usadel equation in the diffusive transport limit. 
It is clear from this analytical formulation that the magnetization in the $N$ is indeed induced 
by the odd-frequency STCs. 
In Sec.~\ref{sec:result}, we show the results of the magnetization as a function of the thickness of $N$ for different 
magnetization alignments of the two $F$s. The $\theta$ dependence of magnetization, including the dynamics when AC voltage is applied, is also discussed. 
Finally, the magnetization induced by the STCs is estimated for a typical set of realistic parameters in Sec.~\ref{sec:discussion}. 
The summary of this paper is given in Sec.~\ref{sec:summary}. 
The spatial dependence of anomalous Green's functions in the $N$ is discussed in 
Appendix~\ref{app:agf} and the local magnetization density induced inside the $N$ is examined in 
Appendix~\ref{app:lmd}.

\section{Junction and formulation}\label{sec:formulation}

After introducing the Josephson junction studied, we first formulate for this junction 
the anomalous Green's functions in the diffusive transport limit on the basis of the quasiclassical 
Green's function method and 
then derive the analytical formulae of the magnetization induced inside the $N$. 

\subsection{S/F1/N/F2/S junction}

\begin{figure}[t!]
\begin{center}
\vspace{10 mm} 
\includegraphics[width=6cm]{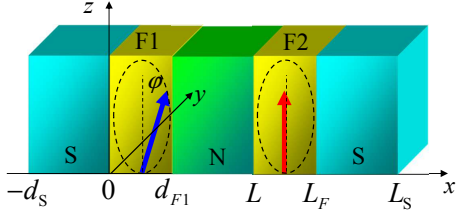}
\caption{ (Color online) 
Schematic illustration of the $S$/$F$1/$N$/$F$2/$S$ junction studied, where the normal metal ($N$) is sandwiched 
by two ferromagnetic metals ($F$1 and $F$2) attached to $s$-wave superconductors ($S$s). 
Arrows in $F$1 and $F$2 indicate the direction of ferromagnetic magnetizations. 
While the magnetization in $F$2 is fixed along the $z$ direction, the $F$1 is assumed to be a free layer 
in which the magnetization can be controlled by an external magnetic field within the $yz$ plane 
with $\varphi $ being the polar angle of the magnetization. 
$d_{\rm S}$, $d_{\rm F1}$, $d_{\rm F2}$, and $d$ are the thicknesses of $S$, $F$1, $F$2, and $N$, 
respectively, with $L=d+d_{\rm F1}$, $L_{\rm F}=L+d_{\rm F2}$, and $L_{\rm S}=L_{\rm F}+d_{\rm S}$. 
We assume that the magnetizations are uniform in both $F$1 and $F$2 layers, and that  
$d_{\rm S}\gg\xi_{\rm S}$.
}
\label{sfnfs-gm}
\end{center}
\end{figure}

As depicted in Fig.~\ref{sfnfs-gm}, we consider the $S$/$F$1/$N$/$F$2/$S$ junction made of normal metal ($N$) 
sandwiched by two layers of ferromagnetic metal ($F$1 and $F$2) 
attached to $s$-wave superconductors ($S$s). 
We assume that the magnetization in $F$2 is fixed along the $z$ direction 
perpendicular to the junction direction ($x$ direction), 
while the $F$1 is a free layer in which the magnetization can be controlled by an external magnetic field, 
pointing any direction in the $yz$ plane, parallel to the interfaces, with 
$\varphi $ being the polar angle of the magnetization. We also assume that 
the magnetizations in $F$1 and $F$2 are both uniform. 
The thicknesses of $S$, $F$1, $F$2, and $N$ are $d_{\rm S}$, $d_{\rm F1}$, $d_{\rm F2}$, and $d$, 
respectively, with $L=d+d_{\rm F1}$, $L_{\rm F}=L+d_{\rm F2}$, and 
$L_{\rm S} = L_{\rm F} + d_{\rm S}$. 
Furthermore, we assume that $d_{\rm S}$ is much larger than the superconducting coherent length 
$\xi_{\rm S}$. 

\subsection{Anomalous Green's functions}\label{sec:agf}

In the diffusive transport region, the magnetization inside the $N$ 
is evaluated by solving the linearized Usadel equation 
in each region $m$ ($=$ $F$1, $N$, and $F$2)~\cite{buzdin-rmp,golubov-rmp,bergeret-rmp}, 
\begin{widetext}
\begin{equation}
i\hbar D_{m}\partial _x^2 {\hat f^{m}(x)} - i2\hbar \left| {\omega _n} \right| {\hat f^{m}(x)} 
+ 2\hat \Delta(x)  - {\mathop{\rm sgn}} \left( {{\omega _n}} \right)h_{\rm ex}^y\left( x \right)\left\{ {{{\hat \tau }_y},{{\hat f}^{m}(x)}} \right\} 
- {\mathop{\rm sgn}} \left( {{\omega _n}} \right)h_{\rm ex}^z\left( x \right)\left[ {{{\hat \tau }_z},{{\hat f}^{m}(x) }} \right] = \hat 0
\label{usadel}, 
\end{equation}
\end{widetext}
where $D_{m}$ is the diffusion coefficient in region $m$, 
$\omega_n=(2n+1)\pi k_{\rm B}T/\hbar$ with $n=0,\pm1,\pm2,\cdots$ is the fermion Matsubara frequency, 
${\rm sgn}(A)=A/|A|$, and $\hat{\tau}_{y(z)}$ is the $y$ ($z$) component of Pauli matrix.
We assume that diffusion coefficients 
in $F$1 and $F$2 are the same, i.e. $D_{\rm F1}=D_{\rm F2}=D_{\rm F}$. 
Note also that $\{\hat{ Q}, \hat{R}\}=\hat{ Q}\hat{R}+\hat{R}\hat{Q}$, $[\hat{Q}, \hat{R}]=\hat{Q}\hat{R}-\hat{R}\hat{Q}$, and $\hat{ 0}$ is null matrix. 
The anomalous part $\hat{f}^{m}$ of the ($2\times2$) quasiclassical Green's function \cite{eschrig-sh} is given by 
\begin{eqnarray}
{\hat f^m}\left( x \right) &=& \left( {\begin{array}{*{20}{c}}
	{f_{ \uparrow  \uparrow }^m\left( x \right)}&{f_{ \uparrow  \downarrow }^m\left( x \right)}\\
	{f_{ \downarrow  \uparrow }^m\left( x \right)}&{f_{ \downarrow  \downarrow }^m\left( x \right)}
	\end{array}} \right) \nonumber \\
	&=& \left( {\begin{array}{*{20}{c}}
	{ - f_{tx}^m\left( x \right) + if_{ty}^m\left( x \right)}&{f_s^m\left( x \right) + f_{tz}^m\left( x \right)}\\
	{ - f_s^m\left( x \right) + f_{tz}^m\left( x \right)}&{f_{tx}^m\left( x \right) + if_{ty}^m\left( x \right)}
\end{array}} \right)
\label{f}, 
\end{eqnarray}
where the $\omega_n$ dependence is implicitly assumed. 
Notice that $f_s^m(x)$ is the anomalous Green's function for the SSC, whereas 
$f_{tx(ty)}^m(x)$ and $f_{tz}^m(x)$ represent the anomalous Green's functions for 
the STC with $|S_z|=1$ and $|S_z|=0$, respectively. 
The $s$-wave superconducting gap $\hat{\Delta}(x)$ is finite only in the $S$ 
and assume to be constant, i.e., 
\begin{eqnarray}
\hat \Delta(x)  = \left\{ \begin{array}{l}
\left( {\begin{array}{*{20}{c}}
0&{ - \Delta_{\rm L} }\\
\Delta_{\rm L} &0
\end{array}} \right), - {d_{\rm S}} < x < 0 \\
\left( {\begin{array}{*{20}{c}}
0&{ - \Delta_{\rm R} }\\
\Delta_{\rm R} & 0
\end{array}} \right), L_{\rm F} < x<L_{\rm S}\\ 
\,\,\,\,\,\,\,\,\,\,\,\,\,\,\,\,\,\,\hat 0,\,\,\,\,\,\,\,\,\,\,\,\,\,\,\,\,\,\,\,\,\,\,\,\,\,\,\,\,\,\,{\rm{other}}
\end{array} \right.\ .
\end{eqnarray}
The exchange field ${\vec h}_{\rm ex}(x)=(h_{\rm ex}^{x}(x),h_{\rm ex}^{y}(x),h_{\rm ex}^{z}(x))$ 
due to the ferromagnetic magnetization in the {\it F}s is described by 
\begin{eqnarray}
{\vec h_{\rm ex}}\left( x \right) = \left\{ \begin{array}{l}
h_{\rm ex}^{y} {\vec e}_y + h_{\rm ex}^{z} {\vec e}_z\,,\,\,\,\,\,\,\,\,\,0 < x < {d_{F1}}\\
{h_{\rm ex2}}{{\vec e}_z},\,\,\,\,\,\,\,\,\,\,\,\,\,\,\,\,\,\,\,\,\,\,\,\,\,\,\,\,\,L < x < {L_F},\\
0,\,\,\,\,\,\,\,\,\,\,\,\,\,\,\,\,\,\,\,\,\,\,\,\,\,\,\,\,\,\,\,\,\,\,\,\,\,\,\,\,\,\,\,\,\,\,\,\,\,\,\,{\rm{other}}
\end{array} \right.
\end{eqnarray}
where $h_{\rm ex}^y=h_{\rm ex1}\sin \varphi$, $h_{\rm ex}^z=h_{\rm ex1}\cos \varphi$  
(see Fig.~\ref{sfnfs-gm}), and  
$\vec{e}_{y(z)}$ is a unit vector in the $y$ ($z$) direction. 
We assume that $h_{\rm ex1}$ and $h_{\rm ex2}$ are both positive.

To obtain the solutions of Eq.~(\ref{usadel}), we impose appropriate boundary 
conditions~\cite{demler-prb}, i.e., 
\begin{eqnarray}
{\left. {{{\hat f}^{{\rm{S}}}(x) }} \right|_{x = 0}} &=& {\left. {{{\hat f }^{{\rm{F1}}}}(x) } \right|_{x = 0}}
\label{bc1}, \\
{\left. {{{\hat f }^{{\rm{F1}}}(x) }} \right|_{x = {d_{{\rm{F1}}}}}} &=& {\left. {{{\hat f}^{\rm{N}}(x)}} \right|_{x = {d_{{\rm{F1}}}}}}
\label{bc2}, \\
{\left. {{{\hat f}^{\rm{N}}(x)}} \right|_{x = L}} &=& {\left. {{{\hat f}^{{\rm{F2}}}(x) }} \right|_{x = L}}
\label{bc3}, \\
{\left. {{{\hat f}^{{\rm{F2}}}(x) }} \right|_{x = {L_{\rm{F}}}}} &=& {\left. {{{\hat f}^{{\rm{S}}}(x) }} \right|_{x = {L_{\rm{F}}}}}
\label{bc4}, \\
{\left. {\partial_x {{\hat f}^{{\rm{F1}}}(x) }} \right|_{x = {d_{{\rm{F1}}}}}} &=& {\left. \frac{1}{{{\gamma _{\rm{F}}}}} {\partial_x {{\hat f}^{\rm{N}}(x) }} \right|_{x = {d_{{\rm{F1}}}}}}
\label{bc5}, 
\end{eqnarray}
and 
\begin{eqnarray}
\frac{1}{{{\gamma _{\rm{F}}}}} {\left. {\partial_x {{\hat f}^{\rm{N}}(x) }}  \right|_{x = L}} &=& {\left. {\partial_x {{\hat f}^{{\rm{F2} }}}(x) } \right|_{x = L}}
\label{bc6}, 
\end{eqnarray}
where $\gamma_{\rm F}=\sigma_{\rm F}/\sigma_{\rm N}$ and 
$\sigma_{\rm F(N)}$ is the conductivity of $F$1 and $F$2 ($N$). 
Moreover, in the present calculation,  
we adopt the rigid boundary condition 
\begin{equation}
\frac{\sigma_{\rm F}}{\sigma_{\rm S}} \ll \frac{\xi_{\rm F1(2)}}{\xi_{\rm S}},
\end{equation} 
where $\sigma_{\rm S}$ is the conductivity of $S$ in the normal state and 
$\xi_{\rm F1(2)}=\sqrt{\hbar D_{\rm F}/h_{\rm ex1(2)} }$~\cite{buzdin-rmp}.  
Assuming that $d_{\rm S}\gg\xi_{\rm S}$, the anomalous Green's function in the $S$s attached to $F$1 and $F$2 
can be approximately given as 
\begin{equation}
{\hat f}_{s}^{\rm S} (x) |_{x=0 (L_{\rm F})} = - {\hat \tau}_{y} 
\frac{\Delta_{\rm L (R)}  } {\sqrt{(\hbar \omega)^{2} + |\Delta_{\rm L (R)}|^{2}}},  
\end{equation} 
where $\Delta_{\rm L(R)} = \Delta e^{i \theta_{\rm L(R)}}$ ($\Delta$: real) and $\theta_{\rm L(R)}$ is the superconducting phase in the left (right) side of $S$s 
(see Fig.~\ref{sfnfs-gm}).

Assuming that $d_{\rm F1}/\xi_{\rm F1}\ll 1$, we can preform the Taylar expansion for 
${\hat f^{\rm F1}}(x)$ as follows~\cite{houzet-prb76, ode-book}:  
\begin{eqnarray}
{\hat f^{\rm F1}}\left( x \right) &\approx& {\hat f^{\rm F1}}\left( {{d_{\rm F1}}} \right) 
	+ \left. \left( {x - {d_{\rm F1}}} \right){\partial _x}{\hat f^{\rm F1}}\left( x \right) \right|_{x=d_{\rm F1}} \nonumber \\
	& + & \frac{{{{\left( {x - {d_{\rm F1}}} \right)}^2}}}{2}\left. \partial _x^2{\hat f^{\rm F1}}\left( x \right)\right|_{x=d_{\rm F1}}
\label{ff1}.
\end{eqnarray}
Using the boundary conditions given in Eqs.~(\ref{bc1}) and~(\ref{bc5}) for Eq.~(\ref{ff1}) and 
substituting Eq.~(\ref{ff1}) into Eq.~(\ref{usadel}), 
${\hat f^{\rm F1}}(x)$ can be approximately expressed as 
\begin{widetext}
\begin{eqnarray}
{\hat f}^{\rm F1}(x) &\approx &
		\frac{x}{\gamma_{\rm F}} \left. \partial _{x} \hat{f}^{\rm N}(x)\right|_{x=d_{\rm F1}}
		+ {\hat f}^{\rm S1}(0)
		+ i {\rm sgn}(\omega_{n}) \frac{h_{\rm ex}^{y} d_{\rm F1}^{2} }{2 \hbar D_{\rm F}}
		\left\{ 
		{\hat \tau_{y}, {\hat f}^{\rm S1}(0)}
		\right\}
		+ i {\rm sgn}(\omega_{n}) \frac{h_{\rm ex}^{z} d_{\rm F1}^{2} }{2 \hbar D_{\rm F}}
		\left[ 
		\hat \tau_{z}, {\hat f}^{\rm S1}(0)
		\right] \nonumber \\
		&-& i {\rm sgn}(\omega_{n}) \frac{(x-d_{\rm F1})^{2}}{2\hbar D_{\rm F}} 
		\left(
		h_{\rm ex}^{y} 
		\left\{ 
		{\hat \tau_{y}, {\hat f}^{\rm S1}(0)}
		\right\}
		+
		\left[ 
		\hat \tau_{z}, {\hat f}^{\rm S1}(0)
		\right]
		\right)
\label{ff1x}.
\end{eqnarray}
\end{widetext}
Here we also assume that the exchange field $h_{{\rm ex}1}$ 
in the $F$1 is much larger than $k_{\rm B}T$ and thus 
the term $\hbar|\omega_{n}|{\hat f^{\rm F1}}(x)$ is neglected in Eq.~(\ref{ff1x}).

Similarly, assuming that $d_{\rm F2}/\xi_{\rm F2}\ll 1$, we can perform the Taylor expansion for 
${\hat f^{\rm F2}}(x)$ and, using the boundary conditions given in Eqs.~(\ref{bc4}) and (\ref{bc6}), 
${\hat f^{\rm F2}}(x)$ can be approximately expressed as 
\begin{widetext}
\begin{eqnarray}
{\hat f}^{\rm F2}(x) &\approx &
		-\frac{d_{\rm F2}}{\gamma_{\rm F}} \left. \partial _{x} \hat{f}^{\rm N}(x)\right|_{x=L} 
		+ {\hat f}^{\rm S2}(L_{\rm F})
		+\frac{x-L}{\gamma_{\rm F}} \left. \partial _{x} \hat{f}^{\rm N}(x)\right|_{x=L} 
		+ i {\rm sgn}(\omega_{n}) \frac{h_{\rm ex2} d_{\rm F2}^{2} }{2 \hbar D_{\rm F}}
		\left[ 
		\hat \tau_{z}, {\hat f}^{\rm S2}(L_{\rm F})
		\right] \nonumber \\
		&-& i {\rm sgn}(\omega_{n}) \frac{(x-L)^{2} h_{\rm ex2}}{2\hbar D_{\rm F}} 
		\left[ 
		\hat \tau_{z}, {\hat f}^{\rm S2}(L_{\rm F})
		\right]
\label{ff2x}, 
\end{eqnarray}
\end{widetext}
where $h_{\rm ex2}\gg k_{\rm B}T$ is also assumed.

The general solutions of ${\hat f}^{\rm N}(x)$ are given as 
\begin{eqnarray}
	f_{\pm }^{\rm N}(x) &=& A_{\pm }^{\rm N} e^{k_{\rm N} x } + B_{\pm}^{\rm N} e^{-k_{\rm N} x}
\label{fnpm1} 
\end{eqnarray}
and 
\begin{eqnarray}
	f_{ty}^{\rm N} (x) &=&  A_{y}^{\rm N} e^{k_{\rm N} x } + B_{y}^{\rm N} e^{-k_{\rm N} x}
\label{fny}, 
\end{eqnarray}
where 
\begin{eqnarray}
	f_{\pm }^{\rm N}(x) &=& f_{s}^{\rm N}(x) \pm  f_{tz}^{\rm N}(x)
\label{fnpm2} 
\end{eqnarray}
and $k_{\rm N}=\sqrt{2|\omega_{n}|/D_{\rm N}}$. 
Applying the boundary conditions given in Eqs.~(\ref{bc2}) and (\ref{bc3}) to Eqs.~(\ref{fnpm1}) 
and (\ref{fny}), and also using the results in Eqs.~(\ref{ff1x}) and (\ref{ff2x}), 
we can obtain the anomalous Green's functions in the $N$ as 
\begin{widetext}
\begin{eqnarray}
	f_{s}^{\rm N} (x) &=& -i \frac{\Delta_{\rm L}}{E_{\omega_{n}}} 
	\left[
	{\rm sinh}[k_{\rm N}(x-L)] -\frac{k_{\rm N}d_{\rm F2}}{\gamma_{\rm F}} {\rm cosh}[k_{\rm N}(x-L)]
	\right] K_{\omega_{n}}(d) \nonumber \\
	&+& i \frac{\Delta_{\rm R}}{E_{\omega_{n}}}
	\left[
	{\rm sinh}[k_{\rm N}(x-d_{\rm F1})] +\frac{k_{\rm N}d_{\rm F1}}{\gamma_{\rm F}} {\rm cosh}[k_{\rm N}(x-d_{\rm F1})]
	\right] K_{\omega_{n}}(d)
\label{fns}, \\
	f_{ty}^{\rm N}(x) &=& {\rm sgn}(\omega_{n}) \frac{\Delta_{\rm L}}{E_{\omega_{n}}} \frac{h_{\rm ex}^{y} d_{\rm F1}^{2}}{\hbar D_{\rm F}}
	\left[
	{\rm sinh}[k_{\rm N}(x-L)] -\frac{k_{\rm N}d_{\rm F2}}{\gamma_{\rm F}} {\rm cosh}[k_{\rm N}(x-L)]
	\right] K_{\omega_{n}}(d)
\label{fnty}, 
\end{eqnarray}
and 
\begin{eqnarray}
	f_{tz}^{\rm N} (x) &=& {\rm sgn}(\omega_{n}) \frac{\Delta_{\rm L}}{E_{\omega_{n}}} \frac{h_{\rm ex}^{z} d_{\rm F1}^{2}}{\hbar D_{\rm F}} 
	\left[
	{\rm sinh}[k_{\rm N}(x-L)] -\frac{k_{\rm N}d_{\rm F2}}{\gamma_{\rm F}} {\rm cosh}[k_{\rm N}(x-L)]
	\right] K_{\omega_{n}}(d) \nonumber \\
	&-& {\rm sgn}(\omega_{n}) \frac{\Delta_{\rm R}}{E_{\omega_{n}}} \frac{h_{\rm ex2} d_{\rm F2}^{2}}{\hbar D_{\rm F}} 
	\left[
	{\rm sinh}[k_{\rm N}(x-d_{\rm F1})] +\frac{k_{\rm N}d_{\rm F1}}{\gamma_{\rm F}} {\rm cosh}[k_{\rm N}(x-d_{\rm F1})]
	\right] K_{\omega_{n}}(d)
\label{fntz}, 
\end{eqnarray}
\end{widetext}
where 
\begin{eqnarray}
E_{\omega_n} &=& \sqrt{(\hbar \omega_{n})^{2} + \Delta^{2}} 
\end{eqnarray}
and 
\begin{eqnarray}
	K_{\omega_{n}}^{-1}(d) &=& 
	\left(
	\frac{k_{\rm N} d_{\rm F1}}{\gamma_{\rm F}} +\frac{k_{\rm N} d_{\rm F2}}{\gamma_{\rm F}} 
	\right)
	{\rm cosh}(k_{\rm N} d) \nonumber \\
	&+&
	\left(
	1+\frac{k_{\rm N} d_{\rm F1}}{\gamma_{\rm F}} \frac{k_{\rm N} d_{\rm F2}}{\gamma_{\rm F}}
	\right) 
	{\rm sinh}(k_{\rm N} d)
\label{kwd}. 
\end{eqnarray}
From Eqs.~(\ref{fns})--(\ref{fntz}), it is immediately found that that $f_{s}^{\rm N}(x)$ describing 
the SSC is an even function with respect to $\omega_{n}$, whereas 
$f_{ty(tz)}^{\rm N}(x)$ describing the STC is an odd function with respect to 
$\omega_{n}$ since $f_{ty(tz)}^{\rm N}(x)$ is proportional to ${\rm sgn}(\omega_{n})$.
Hence, $f_{ty(tz)}^{\rm N}(x)$ represents the odd-frequency STC. 

It should be emphasized here that 
\begin{equation}
\lim_{h_{{\rm ex}1}\to0}f_{ty}^{\rm N}(x) = 0
\end{equation}
and
\begin{equation}
\lim_{h_{{\rm ex}1},h_{{\rm ex}2}\to0}f_{tz}^{\rm N}(x) = 0,
\end{equation}
whereas $f_{s}^{\rm N} (x)$ is generally finite independently of $h_{{\rm ex}1}$ and $h_{{\rm ex}2}$. 
This is due to the fact that the presence of $F$ layers are essential to induce the STC~\cite{bergeret-rmp}. 
On the contrary, the SSC is 
always induced inside the $N$ in $S$/$N$ junctions as well as more complex 
$S$/$N$/$F$ junctions~\cite{yokoyama-prb80}. 
Notice also that i) $f_{tx}^{\rm N}(x)=0$ because the exchange field in the $F$1 does not 
have the $x$ component. 
and ii) $f_{ty}^{\rm N}(x)$ is exactly zero when $\varphi =0$ or 
$\pi$ as $f_{ty}^{\rm N}(x) \propto h_{\rm ex}^{y}$. 
The spacial dependence of anomalous Green's functions in the $N$ is discussed 
in Appendix~\ref{app:agf}.

\subsection{Induced magnetization in normal metal}

Within the quasiclassical Green's function method, 
the magnetization $\vec{M}(d,\theta)$ 
induced inside the $N$ is given~\cite{Lofwander-prl95,champel-prb72} as 
\begin{eqnarray}
	\vec{M}(d,\theta) &=&
		(M_{x}(d,\theta),M_{y}(d,\theta),M_{z}(d,\theta)) \nonumber \\
		&=&
		\frac{A}{V}
		\int_{d_{F1}}^{L}
		\vec{m}(x,\theta) dx
\label{md},
\end{eqnarray}
where $\theta=\theta_{\rm R}-\theta_{\rm L}$ is the superconducting phase difference between the outmost $S$s in the junction and 
\begin{eqnarray}
	\vec{m} (x,\theta) &=& (m_{x}(x,\theta),m_{y}(x,\theta),m_{z}(x,\theta)) \nonumber \\
	&=&- g \mu_{\rm B} \pi N_{\rm F} k_{\rm B} T 
	\sum_{\omega_{n}}
	{\rm sgn}(\omega_{n})
	{\rm Im}
	\left[
	f_{s}^{\rm N}(x) \vec{f}_{t}^{\,\,\rm N*}(x) 
	\right] \nonumber \\
\label{m}
\end{eqnarray}
with 
\begin{equation}
{\vec{f}_{t}}^{\,\,\rm N}(x) = 
	(f_{tx}^{\rm N}(x),-f_{ty}^{\rm N}(x),f_{tz}^{\rm N}(x))
\label{vec-f}.
\end{equation}
Here, $\vec m(x,\theta)$ is the local magnetization density 
in the $N$, $g$ is the $g$ factor of electron, $\mu_{\rm B}$ is the Bohr magneton, and $A$ and $V=Ad$ 
are the cross-section area of junction and the volume of $N$, respectively. 
In the quasiclassical Green's function method, the density of states $N_{\rm F}$ per unit volume 
and per electron spin at the Fermi energy 
is assumed to be approximately the same
for up and down electrons in the $N$~\cite{golubov-rmp,buzdin-rmp,bergeret-rmp}.

It is apparent in Eq.~(\ref{m}) that $f_{s}^{\rm N}(x)$ and $\vec{f}_{t}^{\,\,\rm N}(x)$ are both required to be nonzero to induce finite 
$\vec{m} (x,\theta)$. However, as described in Sec.~\ref{sec:agf}, nonzero $\vec{f}_{t}^{\,\,\rm N}(x)$ 
occurs only when $F$ layers are involved in the junction and $\vec{f}_{t}^{\,\,\rm N}(x)=0$ whenever 
$f_{s}^{\rm N}(x)=0$ for $\Delta_{\rm L}=\Delta_{\rm R}=0$. Therefore, the origin of the magnetization 
in the $N$ is considered to be due to the STCs induced by the proximity effect~\cite{bergeret-rmp,Lofwander-prl95,pugach-apl101}.  
Note also that because of $f_{tx}^{\rm N}(x)=0$ (see Sec.~\ref{sec:agf}), 
$m_{x}(x,\theta)$ and thus $M_{x} (d,\theta)$ are always zero. 
Therefore, in the following, we only consider the $y$ and $z$ components of $\vec{M}(d,\theta)$. 
More details of $\vec m(x,\theta)$ are examined in Appendix~\ref{app:lmd}. 

Substituting Eqs~(\ref{fns})--(\ref{fntz}) into Eq.~(\ref{m}) and 
performing the integration with respect to $x$ in Eq.~(\ref{md}), 
we can obtain the $y$ and $z$ components of the magnetization induced 
inside the $N$. 
The $y$ component $M_{y}(d,\theta)$ of the magnetization is decomposed into two parts,  
\begin{eqnarray}
	M_{y}(d,\theta) &=& M_{y}^{\rm I}(d) + M_{y}^{\rm II}(d,\theta)
\label{my}, 
\end{eqnarray}
where 
\begin{widetext}
\begin{eqnarray}
	M_{y}^{\rm I}(d) &=& 
	-g \mu_{\rm B} k_{\rm B} T \frac{\pi N_{\rm F} \Delta^{2}}{2 d} 
	\frac{h_{\rm ex}^{y} d_{\rm F1}^{2}}{\hbar D_{\rm F}} 
	\sum_{\omega_{n}}
	\frac{ K_{\omega_{n}}^{2}(d) F_{\omega_{n}}^{\rm I}(d) }{ k_{\rm N} E_{\omega_{n}}^{2} }
\label{my1}
\end{eqnarray}
and 
\begin{eqnarray}
	M_{y}^{\rm II}(d,\theta) &=&
	g \mu_{\rm B} k_{\rm B} T \frac{\pi N_{\rm F} \Delta^{2}}{2 d} 
	\frac{h_{\rm ex}^{y} d_{\rm F1}^{2}}{\hbar D_{\rm F}} 
	\sum_{\omega_{n}}
	\frac{ K_{\omega_{n}}^{2}(d) F_{\omega_{n}}^{\rm II}(d) }{ k_{\rm N} E_{\omega_{n}}^{2} } {\rm cos}\theta
\label{my2}. 
\end{eqnarray} 
Here, we have introduced 
\begin{eqnarray}
	F_{\omega_{n}}^{\rm I}(d) &=& \frac{k_{\rm N}d_{\rm F2}}{\gamma_{\rm F}}
	\left[
	1-{\rm cos}(2k_{\rm N} d) 
	\right]
	+
	\left[
	1-\left(\frac{k_{\rm N} d_{\rm F2} }{\gamma_{\rm F}} \right)^{2} 
	\right] k_{\rm N} d
	-\frac{1}{2}
	\left[
	1+\left(\frac{k_{\rm N} d_{\rm F2} }{\gamma_{\rm F}} \right)^{2} 
	\right]  {\rm sinh}(2k_{\rm N} d) 
\end{eqnarray}
and 
\begin{eqnarray}
	F_{\omega_{n}}^{\rm II}(d) &=&
	\left(
	1+\frac{ k_{\rm N}d_{\rm F1} }{\gamma_{\rm F}} \frac{ k_{\rm N}d_{\rm F2} }{\gamma_{\rm F}}
	\right) k_{\rm N} d
	{\rm cosh}(k_{\rm N}d)
	-
	\left[
	1
	-\frac{ k_{\rm N}d_{\rm F1} }{\gamma_{\rm F}} \frac{ k_{\rm N}d_{\rm F2} }{\gamma_{\rm F}}
	-
	\left(
	\frac{ k_{\rm N}d_{\rm F1} }{\gamma_{\rm F}} + \frac{ k_{\rm N}d_{\rm F2} }{\gamma_{\rm F}}
	\right) k_{\rm N} d
	\right] {\rm sinh}(k_{\rm N} d).
\end{eqnarray}
\end{widetext}
Similarly, the $z$ component $M_{z}(d,\theta)$ of the magnetization is decomposed into two parts,  
\begin{eqnarray}
	M_{z}(d,\theta) &=& M_{z}^{\rm I}(d) + M_{z}^{\rm II}(d,\theta)
\label{mz}, 
\end{eqnarray}
where
\begin{widetext}
\begin{eqnarray}
	M_{z}^{\rm I}(d) &=& g \mu_{\rm B} k_{\rm B} T \frac{ \pi N_{\rm F} \Delta^{2} }{ 2d }
		\sum_{\omega_{n}}
		\frac{ K_{\omega_{n}}^{2}(d) }{ k_{\rm N} E_{\omega_{n}}^{2} }
		\left[
		\frac{ h_{\rm ex}^{z} d_{\rm F1}^{2} }{ \hbar D_{\rm F} } R_{\omega_{n}}^{\rm a}(d) 
		+ \frac{ h_{\rm ex2} d_{\rm F2}^{2} }{ \hbar D_{\rm F} } R_{\omega_{n}}^{\rm b}(d) 
		\right]
\label{mz1} 
\end{eqnarray}
and
\begin{eqnarray}
	M_{z}^{\rm II}(d,\theta) &=& -g \mu_{\rm B} k_{\rm B} T \frac{ \pi N_{\rm F} \Delta^{2} }{ 2d }
	\left(
		\frac{h_{\rm ex}^{z} d_{\rm F1}^{2}}{ \hbar D_{\rm F} } 
		+ \frac{h_{\rm ex2} d_{\rm F2}^{2}}{ \hbar D_{\rm F} }
	\right)
	\sum_{\omega_{n}}
	\frac{K_{\omega_{n}}^{2}(d)F_{\omega_{n}}^{\rm II}(d) }{ k_{\rm N} E_{\omega_{n}}^{2} }
	{\rm cos}\theta
\label{mz2}. 
\end{eqnarray}
Here, we have also introduced 
\begin{eqnarray}
	R_{\omega_{n}}^{\rm a(b)}(d) &=&
		\frac{k_{\rm N} d_{\rm F1(2)}}{\gamma_{\rm F}}
		+
		\left[
		1-\left( \frac{k_{\rm N} d_{\rm F1(2)}}{\gamma_{\rm F}} \right)^{2}
		\right] k_{\rm N} d
		-
		\frac{k_{\rm N} d_{\rm F1(2)}}{\gamma_{\rm F}}
		{\rm cosh}(2k_{\rm N} d)
		- \frac{1}{2}
		\left[
		1+\left( \frac{k_{\rm N} d_{\rm F1(2)}}{\gamma_{\rm F}} \right)^{2}
		\right]
		{\rm sinh}(2k_{\rm N} d)
		\label{eq:rI}.
\end{eqnarray}
\end{widetext}

The $\theta$ independent part of the magnetization, i.e., $M_{y}^{\rm I}(d)$ and $M_{z}^{\rm I}(d)$, 
is due to the proximity effect common in $S$/$F$ junctions, 
similar to the one inducing the STCs in $F$/$S$/$F$ and $S$/$F$/$F$ 
junctions~\cite{fominov-jetpl77, Lofwander-prl95, halterman-prb77, fominov-jetpl91, kawabata-jpsj82, mironov-prb89}. 
On the other hand, the $\theta$ dependent part $M_{y}^{\rm II}(d,\theta)$ and 
$M_{z}^{\rm II}(d,\theta)$ of the magnetization is induced by the coupling between the two $S$s in the junction. 
Therefore, $M_{y(z)}^{\rm II}(d,\theta)$ appears only when 
ferromagnetic metallic multilayers constitute the Josephson junction~\cite{pugach-apl101}. 
It should also be noticed that $M_{y}(d,\theta)$ becomes zero when $\varphi =0$ or 
$\pi$ since $M_{y}^{\rm I}(d)$ and $M_{\rm II}(d,\theta)$ are both proportional to 
$h_{\rm ex}^{y}=h_{\rm ex1}\sin \varphi$. 
In contrast, $M_{z}(d,\theta)$ is generally nonzero for any $\varphi$.

\section{Results}\label{sec:result}

\subsection{Thickness dependence of magnetization in normal metal} 

\begin{figure}
\begin{center}
\vspace{10mm} 
\includegraphics[width=6.5cm]{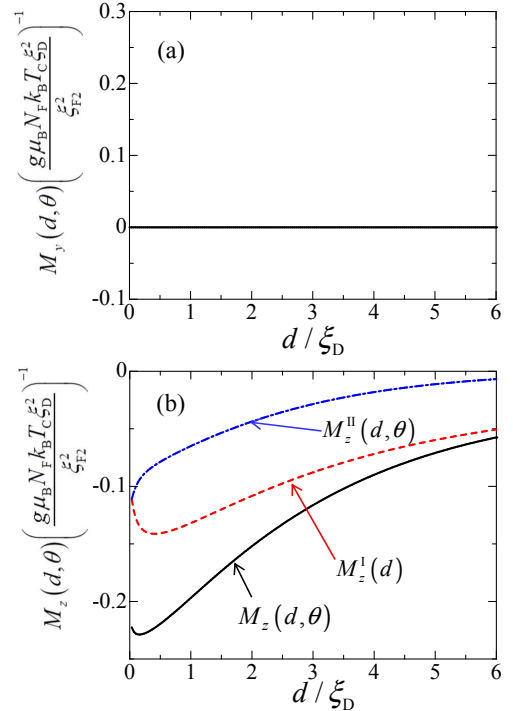}
\caption{ (Color online) (a) The $y$ component $M_{y}(d,\theta)$ and 
(b) the $z$ component $M_{z}(d,\theta)$ of the magnetization in the $N$ 
for $\varphi=0$, corresponding to the parallel magnetization configuration between 
$F$1 and $F$2. 
For other parameters, we set $T/T_{\rm C} = 0.3$,  $\theta=0$, $\gamma_{\rm F} = 0.1$, 
$d_{\rm F1}/\xi_{\rm D} = 0.3$, $d_{\rm F2}/\xi_{\rm D} = 0.2$, $h_{\rm ex1}/\Delta_{0}=30$, 
and $h_{\rm ex2}/\Delta_{0}=20$. 
For comparison, $M_{z}^{\rm I}(d)$ and $M_{z}^{\rm II}(d,\theta)$ are also plotted separately in (b). 
}
\label{m1}
\end{center}
\end{figure}

\begin{figure}[t!]
\begin{center}
\vspace{10mm} 
\includegraphics[width=6.5cm]{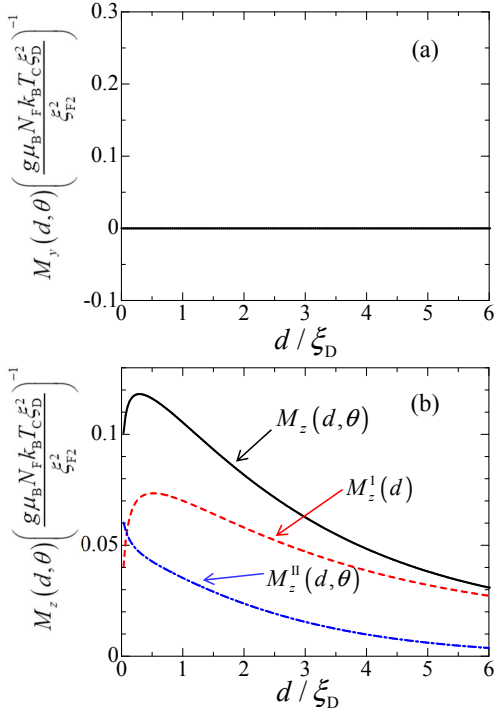}
\caption{ (Color online) (a) The $y$ component $M_{y}(d,\theta)$ and (b) the $z$ component 
$M_{z}(d,\theta)$ of the magnetization in the $N$ for $\varphi =\pi$, 
corresponding to the antiparallel magnetization configuration 
between $F$1 and $F$2. 
Other parameters are the same as in Fig.~\ref{m1}. 
For comparison, $M_{z}^{\rm I}(d)$ and $M_{z}^{\rm II}(d,\theta)$ are also plotted separately in (b). 
}
\label{m2}
\end{center}
\end{figure}

\begin{figure}[t!]
\begin{center}
\vspace{10mm} 
\includegraphics[width=6.5cm]{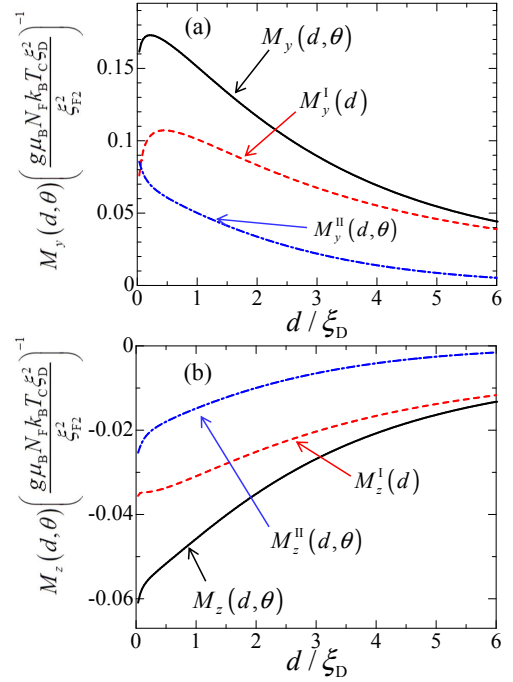}
\caption{ (Color online) (a) The $y$ component $M_{y}(d,\theta)$ and (b) the $z$ component 
$M_{z}(d,\theta)$ of the magnetization in the $N$ for $\varphi =\pi/2$, 
corresponding to the case where the magnetization in $F$1 is perpendicular to that in $F$2. 
Other parameters are the same as in Fig.~\ref{m1}. 
For comparison, $M_{y(z)}^{\rm I}(d)$ and $M_{y(z)}^{\rm II}(d,\theta)$ are also plotted separately.  
}
\label{m3}
\end{center}
\end{figure}

Let us first numerically evaluate $M_{y}(d,\theta)$ and $M_{z}(d,\theta)$ in the $N$ obtained in 
Eqs.~(\ref{my})--(\ref{eq:rI}). 
For this purpose, 
the temperature dependence of $\Delta$ is assumed as 
\begin{equation}
\Delta=\Delta_{0}\tanh\left(1.74\sqrt{\frac{T_{\rm C}}{T}-1}\right), 
\end{equation} 
where $\Delta_{0}$ is the superconducting gap at zero temperature and $T_{\rm C}$ 
is the superconducting transition temperature~\cite{book-sc}. 
Figures~\ref{m1}--\ref{m3} show the typical results of the magnetization in the $N$ as a function of 
thickness $d$ of the $N$ normalized by $\xi_{\rm D}=\sqrt{\hbar D_{\rm N}/2\pi k_{\rm B} T_{\rm C}}$.

Figure~\ref{m1} represents the results for $\varphi = 0$ where the magnetizations between 
F1 and F2 are parallel. 
As shown in Fig.~\ref{m1}(a), the $y$ component $M_{y}(d,\theta)$ of the magnetization is exactly zero 
since $f_{ty}^{\rm N}(x)$ contributing to $M_{y}(d,\theta)$ is zero in the parallel magnetization 
configuration. 
On the other hand, the $z$ component $M_{z}(d,\theta)$ has a finite value, 
as shown in Fig.~\ref{m1}(b), because $f_{tz}^{\rm N}(x)$ contributing to $M_{z}(d,\theta)$ is 
non-zero in the parallel magnetization configuration. 
Furthermore, the induced magnetization $M_{z}(d,\theta)$ is found to be negative, i.e., opposite to 
the magnetizations in F1 and F2. 
It is also found in Fig.~\ref{m1}(b) that $|M_{z}(d,\theta)|$ monotonically decreases with increasing 
$d$ for $d > \xi_{\rm D}$,  
but $M_{z}^{\rm I}(d)$ decays rather slowly as compared 
with $M_{z}^{\rm II}(d,\theta)$. 
The difference of the decay rates for $d\gg\xi_{\rm D}$ as well as the small $d$ behavior in $M_{z}^{\rm I}(d)$ and $M_{z}^{\rm II}(d,\theta)$ will be 
further discussed below.

Figure~\ref{m2} shows the results for $\varphi = \pi$ where the magnetizations between $F$1 and $F$2 are antiparallel. 
As shown in Fig.~\ref{m2}(a), the $y$ component $M_{y}(d,\theta)$ of the magnetization is still exactly zero 
since $f_{ty}^{\rm N}(x)$ contributing to $M_{y}(d,\theta)$ is zero also in the antiparallel 
magnetization configuration. However, the $z$ component $M_{z}(d,\theta)$ is finite and 
decreases monotonically with increasing $d$ for $d>\xi_{\rm D}$ [see Fig.~\ref{m2}(b)]. 
It is also noticed in Fig.~\ref{m2}(b) that the sign of $M_{z}(d,\theta)$ is positive and is opposite 
to the one for $\varphi = 0$ [Fig.~\ref{m1}(b)]. 
The sign reversal of $M_{z}(d,\theta)$ will be further discussed below. 
It is also observed in Fig.~\ref{m2}(b) that $M_{z}^{\rm I}(d)$ decays slowly with respect to $d$ as compared with 
$M_{z}^{\rm II}(d,\theta)$, similarly to the case when the magnetizations in $F$1 and $F$2 are parallel. 
It should be noted however that $M_{z}(d,\theta)$ becomes exactly 
zero in the antiparallel magnetization configuration 
when $d_{\rm F1}=d_{\rm F2}$ and $|h_{\rm ex}^{z}|$ $= |h_{\rm ex2}|$ because in this case 
$M_z^{\rm I}(d)$ and $M_z^{\rm II}(d,\theta)$ are both zero, as seen in Eqs.~(\ref{mz1})--(\ref{eq:rI}) 
[see also Eqs.~(\ref{ap-mz1})--(\ref{hex})].

Figure~\ref{m3} shows the results for $\varphi = \pi/2$ 
where the magnetization in $F$1 is perpendicular to that in $F$2. 
As shown in Fig.~\ref{m3}(a), the $y$ component $M_{y}(d,\theta)$ of the magnetization 
is now finite since $f_{ty}^{\rm N}(x)$ contributing to $M_{y}(d,\theta)$ is nonzero in this case. 
It is also found in Fig.~\ref{m3} that both $|M_{y}(d,\theta)|$ and $|M_{z}(d,\theta)|$ decrease monotonically with 
increasing $d$ for $d>\xi_{\rm D}$. 
Moreover, it is clearly observed that the decay rate of $|M_{y(z)}^{\rm I}(d)|$ with respect to $d$ is 
slower than that of $|M_{y(z)}^{\rm II}(d,\theta)|$. 
This is similar to the other cases discussed above in Figs.~\ref{m1} and \ref{m2}.

It is now instructive to consider limiting cases for the magnetization $M_{y(z)}^{\rm I}(d)$ and 
$M_{y(z)}^{\rm II}(d,\theta)$ induced inside the $N$ and analyze the qualitative behavior with 
respect to the thickness $d$ of the $N$. 
For $T\approx T_{\rm C}$ and $d \gg \xi_{\rm D}$, the $y$ components 
$M_{y}^{\rm I}(d)$ and $M_{y}^{\rm II}(d,\theta)$ of the magnetization 
are approximately given as 
\begin{eqnarray}
M_{y}^{\rm I}(d) &\approx & 
	\frac{ g \mu_{\rm B} N_{\rm F} \Delta^{2} }{2\pi k_{\rm B} T}
	\frac{ h_{\rm ex}^{y} d_{\rm F1}^{2} }{\hbar D_{\rm F}} 
	\frac{\xi_{\rm N}}{d}
\label{ap-my1} 
\end{eqnarray}
and
\begin{eqnarray}
M_{y}^{\rm II}(d,\theta) &\approx &
	\frac{ g \mu_{\rm B} N_{\rm F} \Delta^{2} }{\pi k_{\rm B} T}
	\frac{ h_{\rm ex}^{y} d_{\rm F1}^{2} }{\hbar D_{\rm F}} {\rm cos}\theta
	e^{-d/\xi_{\rm N}}
\label{ap-my2}, 
\end{eqnarray}
whereas the $z$ components 
$M_{z}^{\rm I}(d)$ and $M_{z}^{\rm II}(d,\theta)$ of the magnetization 
are approximately 
\begin{eqnarray}
M_{z}^{\rm I}(d) & \approx &
	-\frac{ g \mu_{\rm B} N_{\rm F} \Delta^{2} }{2\pi k_{\rm B} T} 
	\frac{H_{\rm ex}}{\hbar D_{\rm F}}
	\frac{\xi_{\rm N}}{d}
\label{ap-mz1}
\end{eqnarray}
and
\begin{eqnarray}
M_{z}^{\rm II}(d,\theta) & \approx &
	-\frac{ g \mu_{\rm B} N_{\rm F} \Delta^{2} }{\pi k_{\rm B} T} 
	\frac{H_{\rm ex}}{\hbar D_{\rm F}} \cos \theta
	e^{-d/\xi_{\rm N}}
\label{ap-mz2}, 
\end{eqnarray}
where
\begin{eqnarray}
H_{\rm ex} &=& 
	h_{\rm ex}^{z} d_{\rm F1}^{2} 
	+
	h_{\rm ex2} d_{\rm F2}^{2} 
\label{hex} 
\end{eqnarray}
and $\xi_{\rm N}=\sqrt{\hbar D_{\rm N}/2 \pi k_{\rm B} T}$ 
($\approx\xi_{\rm D}$ at $T\approx T_{\rm C}$). 
It is immediately found in Eqs.~(\ref{ap-my1})--(\ref{ap-mz2}) that 
$M_{y(z)}^{\rm I}(d)$ decreases rather slowly, i.e., algebraically, as $1/d$, whereas 
$M_{y(z)}^{\rm II}(d,\theta)$ decays exponentially. 
This is indeed comparable with the numerical results shown in Figs.~\ref{m1}--\ref{m3}. 

Next, we discuss in the same limiting case the sign change of $M_{z}(d,\theta)$ by flipping the magnetization direction 
from the parallel to the antiparallel configuration in $F$1 and $F$2. 
For this purpose, we focus on $H_{\rm ex}$ appearing in Eqs.~(\ref{ap-mz1}) 
and (\ref{ap-mz2}), the definition being given in Eq.~(\ref{hex}). 
In the case of parallel magnetization configuration, $M_{z}(d,\theta)$ is always negative, 
as shown in Fig.~\ref{m1}, simply 
because $H_{\rm ex}$ is positive (assuming that $|\theta|\le\pi/2$). 
On the other hand, in the case of antiparallel magnetization configuration, 
$h_{\rm ex}^{z}$ is negative. Therefore, $M_{z}(d,\theta)$ becomes positive when 
$|h_{\rm ex}^{z}| d_{\rm F1}^{2}$ is larger than $h_{\rm ex2} d_{\rm F2}^{2}$, 
as shown in Fig.~\ref{m2}.

Let us now consider the opposite limit, i.e., $d \ll  \xi_{\rm D}$, at $T\approx T_{\rm C}$. 
In this limit, the $y$ components 
$M_{y}^{\rm I}(d)$ and $M_{y}^{\rm II}(d,\theta)$ of the magnetization are approximately given as 
\begin{eqnarray}
	M_{y}^{\rm I}(d) & \approx &
		\frac{ g \mu_{\rm B} N_{\rm F} \Delta^{2} }{\pi k_{\rm B} T}
		\frac{ h_{\rm ex}^{y} d_{\rm F1}^{2} }{\hbar D_{\rm F}} 
		\left(
		\frac{d_{\rm F1}}{\xi_{\rm N} \gamma_{\rm F}}
		+
		\frac{d_{\rm F2}}{\xi_{\rm N} \gamma_{\rm F}}
		\right)^{-2} \nonumber \\
		&\times&
		\frac{d_{\rm F2}}{\xi_{\rm F} \gamma_{\rm F}}
		\left(
		\frac{d_{\rm F2}}{\xi_{\rm N} \gamma_{\rm F}}
		+
		\frac{d}{\xi_{\rm N}}
		\right)
\label{my-ap2} 
\end{eqnarray}
and
\begin{eqnarray}
	M_{y}^{\rm II}(d,\theta) & \approx & 
		\frac{g \mu_{\rm B} N_{\rm F} \Delta^{2}}{ \pi k_{\rm B} T }
		\frac{h_{\rm ex}^{y} d_{\rm F1}^{2} }{\hbar D_{\rm F}}
		\left(
		\frac{d_{\rm F1}}{\xi_{\rm N} \gamma_{\rm F} }
		+
		\frac{d_{\rm F2}}{\xi_{\rm N} \gamma_{\rm F}}
		\right)^{-2} \nonumber \\
		&\times&
		\left[
		\frac{d_{\rm F1}}{\xi_{\rm N} \gamma_{\rm F}}
		\frac{d_{\rm F2}}{\xi_{\rm N} \gamma_{\rm F}}
		+
		\left(
		\frac{d_{\rm F1}}{2\xi_{\rm N} \gamma_{\rm F}}
		+
		\frac{d_{\rm F2}}{2\xi_{\rm N} \gamma_{\rm F}}		
		\right)
		\frac{d}{\xi_{\rm N}}
		\right]\cos \theta, \nonumber \\
\label{my-ap2-II}
\end{eqnarray}
whereas the $z$ components $M_{z}^{\rm I}(d)$ and $M_{z}^{\rm II}(d,\theta)$ of the magnetization 
are approximately 
\begin{eqnarray}
	M_{z}^{\rm I}(d) & \approx & -
		\frac{ g \mu_{\rm B} N_{\rm F} \Delta^{2} }{\pi k_{\rm B} T}
		\frac{ h_{\rm ex2} d_{\rm F1}^{2} }{\hbar D_{\rm F}} 
		\left(
		\frac{d_{\rm F1}}{\xi_{\rm N} \gamma_{\rm F}}
		+
		\frac{d_{\rm F2}}{\xi_{\rm N} \gamma_{\rm F}}
		\right)^{-2} \nonumber \\
		&\times&
		\left[
		\frac{h_{\rm ex}^{z}}{h_{\rm ex2}}
		\left(
		\frac{d_{\rm F2}}{\xi_{\rm N} \gamma_{\rm F}}
		+
		\frac{d}{\xi_{\rm N}}
		\right)
		+
		\left(
		\frac{d_{\rm F2}}{\xi_{\rm N} \gamma_{\rm F}}
		\right)^{2}
		\left(
		1 + \gamma_{\rm F} \frac{d}{d_{\rm F1}}
		\right)
		\right]
\label{mz-ap2} \nonumber \\
\end{eqnarray}
and 
\begin{widetext}
\begin{eqnarray}
	M_{z}^{\rm II}(d,\theta) & \approx & 
		-\frac{g \mu_{\rm B} N_{\rm F} \Delta^{2}}{ \pi k_{\rm B} T }
		\left(
		\frac{h_{\rm ex}^{z} d_{\rm F1}^{2}}{ \hbar D_{\rm F} }
		+
		\frac{h_{\rm ex2} d_{\rm F2}^{2}}{ \hbar D_{\rm F} }
		\right)
		\left(
		\frac{d_{\rm F1}}{\xi_{\rm N} \gamma_{\rm F} }
		+
		\frac{d_{\rm F2}}{\xi_{\rm N} \gamma_{\rm F}}
		\right)^{-2} 
		\left[
		\frac{d_{\rm F1}}{\xi_{\rm N} \gamma_{\rm F}}
		\frac{d_{\rm F2}}{\xi_{\rm N} \gamma_{\rm F}}
		+
		\left(
		\frac{d_{\rm F1}}{2\xi_{\rm N} \gamma_{\rm F}}
		+
		\frac{d_{\rm F2}}{2\xi_{\rm N} \gamma_{\rm F}}		
		\right)
		\frac{d}{\xi_{\rm N}}
		\right]\cos \theta.  \nonumber \\
\label{mz-ap2-II}
\end{eqnarray}
\end{widetext}
It is therefore readily noticed in Eqs.~(\ref{my-ap2})--(\ref{mz-ap2-II}) that 
$M_{y(z)}^{\rm I}(d)$ and $M_{y(z)}^{\rm II}(d,\theta)$ are linearly dependent on $d$ and their slopes 
are determined by the signs of $h_{\rm ex}^y$ and $h_{\rm ex}^z$. 
This is in good qualitative agreement with the numerical results shown in 
Figs.~\ref{m1}--\ref{m3}. 

\begin{figure*}[t!]
\begin{center}
\includegraphics[width=17cm]{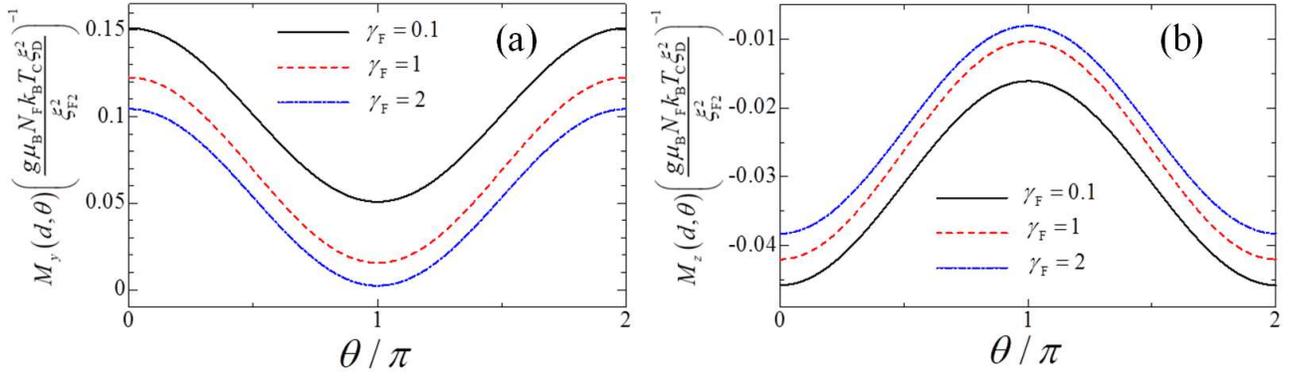}
\caption{ (Color online) 
The $\theta$ dependence of the magnetizations, (a) $M_{y}(d,\theta)$ and (b) $M_{z}(d,\theta)$, 
induced inside the $N$ 
for $\varphi =\pi/2$, corresponding to the case where the magnetization in $F$1 is 
perpendicular to that in $F$2. 
We set $d/\xi_{\rm D} = 1$ for three different values of $\gamma_{\rm F}$ indicated 
in the figures. Other parameters are the same as in Fig.~\ref{m1}. 
}
\label{mq}
\end{center}
\end{figure*}

\subsection{$\theta$ dependence of magnetization in normal metal}

In the previous section, we have focused on the $d$ dependence of the magnetization 
induced inside the $N$. 
Here, we shall demonstrate that the magnetization can also be controlled by the superconducting 
phase difference $\theta$ in the two $S$s. 
The most simplest way to tune $\theta$ experimentally 
is to apply DC bias current to the junction, in which the DC Josephson effect can be detected~\cite{barone}.

Figure~\ref{mq} shows the $y$ and $z$ components $M_y(d,\theta)$ and $M_z(d,\theta)$ of 
the magnetization induced inside the $N$ as a function of $\theta$ for different values of 
$\gamma_{\rm F}$. 
Figure~\ref{mq} clearly demonstrates that the magnetization can indeed be controlled by tuning $\theta$. 
It should also be noticed that the magnitude of the magnetization increases with decreasing 
$\gamma_{\rm F}$. 
Therefore, $\gamma_{\rm F}$ is an important parameter to increase 
the magnetization induced inside the $N$.

\subsection{Dynamics of magnetization in normal metal}

\begin{figure}[t!]
\begin{center}
\vspace{10mm} 
\includegraphics[width=5.5cm]{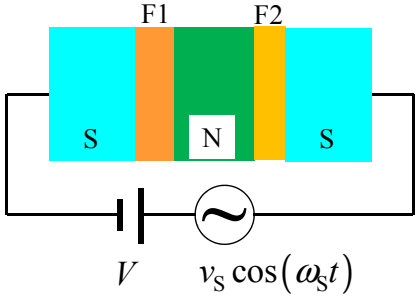}
\caption{ (Color online) 
Schematic set up of the $S$/$F$1/$N$/$F$2/$S$ junction to observe dynamics of the magnetization 
induced inside the $N$. 
This set up is based on the voltage bias model~\cite{barone}, 
where $V$ is the DC voltage, and $v_{\rm S}$ and $\omega_{\rm S}$ are the amplitude and 
frequency of the AC voltage, respectively. 
}
\label{sfnfs-gm2}
\end{center}
\end{figure}

Next, let us discuss an alternative way to control the magnetization induced inside the $N$. 
Here, we consider the $S$/$F$1/$N$/$F$2/$S$ junction subject to both DC and AC external 
fields, described by the voltage bias model~\cite{barone}, as schematically shown in Fig.~\ref{sfnfs-gm2}. 
In this case, the superconducting phase difference $\theta$ evolves in time $t$ according to the 
following well known formula: 
\begin{equation}
	\theta(t) = \theta_{0} + \frac{2e V t}{\hbar} + \frac{2e v_{\rm S} }{\hbar \omega_{\rm S}} \sin(\omega_{\rm S} t)
\label{theta-t}, 
\end{equation}
where $\theta_{0}$ is a time independent constant, $V$ is the DC voltage, 
and $v_{\rm S}$ and $\omega_{\rm S}$ are the amplitude and frequency of the AC voltage, 
respectively.

Substituting Eq.~(\ref{theta-t}) into Eqs.~(\ref{my2}) and (\ref{mz2}), 
and using the generating function of Bessel functions,  
we can easily find that the $\theta$ dependent parts of the 
magnetization in the $N$ are given as 
\begin{widetext}
\begin{eqnarray}
	M_{y}^{\rm II}(d,V,t) &=&
	g \mu_{\rm B} k_{\rm B} T \frac{\pi N_{\rm F} \Delta^{2}}{2 d} 
	\frac{h_{\rm ex}^{y} d_{\rm F1}^{2}}{\hbar D_{\rm F}} 
	\Gamma(V,t)
	\sum_{\omega_{n}}
	\frac{ K_{\omega_{n}}^{2}(d) F_{\omega_{n}}^{\rm II}(d) }{ k_{\rm N} E_{\omega_{n}}^{2} }
\label{dmy}
\end{eqnarray}
and
\begin{eqnarray}
	M_{z}^{\rm II}(d,V,t) &=& -g \mu_{\rm B} k_{\rm B} T \frac{ \pi N_{\rm F} \Delta^{2} }{ 2d }
	\left(
		\frac{h_{\rm ex}^{z} d_{\rm F1}^{2}}{ \hbar D_{\rm F} } 
		+ \frac{h_{\rm ex2} d_{\rm F2}^{2}}{ \hbar D_{\rm F} }
	\right)
	\Gamma(V,t)
	\sum_{\omega_{n}}
	\frac{K_{\omega_{n}}^{2}(d)F_{\omega_{n}}^{\rm II}(d) }{ k_{\rm N} E_{\omega_{n}}^{2} }
\label{dmz} 
\end{eqnarray}
\end{widetext} 
for the $y$ and $z$ components, respectively, 
where the $V$ and $t$ dependence is explicitly shown in the left hand sides. In the above, we have 
also introduced  
\begin{equation}
\Gamma(V,t) = \sum_{m=-\infty }^{\infty }
	(-1)^{m} J_{m}\left( \frac{2\pi v_{\rm s}}{\Phi_{0}\omega_{\rm s}} \right)
	{\rm cos}\left[\theta_{0} + (\omega_{\rm J} - m \omega_{\rm s}) t \right],
	\label{eq:gamma}
\end{equation}
where $\omega_{\rm J}=2eV/\hbar$ is the Josephson frequency, $\Phi_{0}=h/2e$ is the flux quantum, 
and $J_{m}(x)$ is the Bessel function of the first kind ($m$: integer).

\begin{figure*}[t!]
\begin{center}
\vspace{10mm} 
\includegraphics[width=17cm]{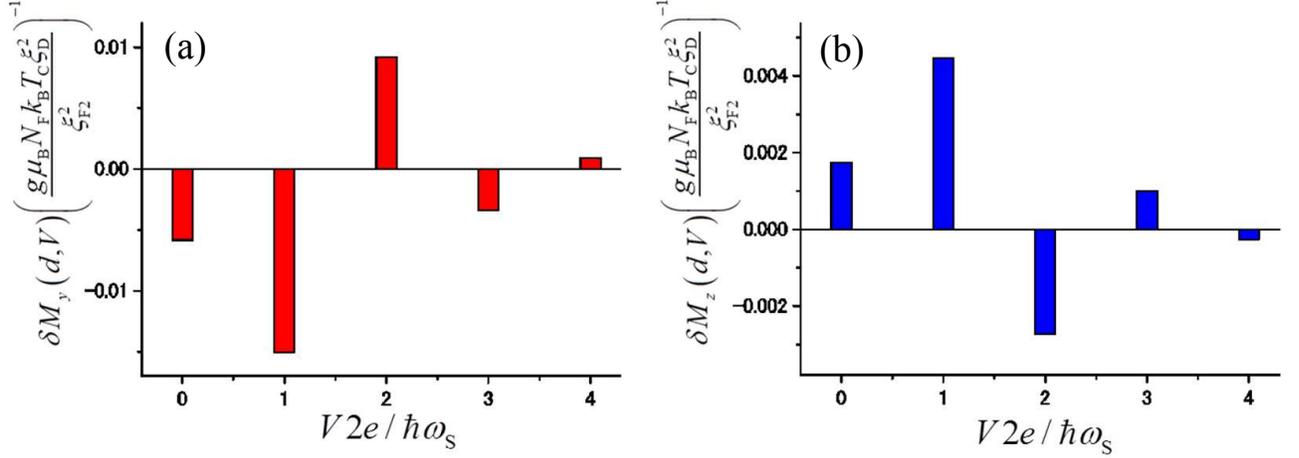}
\caption{ (Color online) 
Time average, (a) $\delta M_{y}(d,V)$ and (b) $\delta M_{z}(d,V)$, for the $\theta$ dependent part of 
the magnetization $M_{y}^{\rm II}(d,V,t)$ and $M_{z}^{\rm II}(d,V,t)$ induced inside the $N$ for 
$\varphi =\pi/2$, corresponding to the case 
where the magnetization in $F$1 is perpendicular to that in $F$2.
We set $d/\xi_{\rm D}=1$, $\theta_{0}=0$, and $\pi v_{\rm S}/\Phi_{0} \omega_{\rm S} =1$. 
Other parameters are the same as in Fig.~\ref{m1}. 
}
\label{mv}
\end{center}
\end{figure*}

Let us now consider the time averaged quantity~\cite{book-time-average} 
\begin{eqnarray}
\delta M_{y(z)}^{\rm II}(d,V) 
& = & \lim_{T \rightarrow \infty }\frac{1}{T}\int_{0}^{T} dt \ M_{y(z)}^{\rm II}(d,V,t). 
\end{eqnarray}
As shown in Eqs.~(\ref{dmy}) and (\ref{dmz}), the $\theta$ dependent part of the magnetization 
clearly oscillates in $t$. Therefore, the time averaged magnetization 
$\delta M_{y(z)}^{\rm II}(d,V)$ is zero except for $\omega_{\rm J}=m \omega_{\rm S}$. 
This is simply because 
the coefficient proportional to $t$ in the cosine function of Eq.~(\ref{eq:gamma}) 
becomes zero only when this condition is satisfied. 
Indeed, the characteristic feature of $\delta M_{y(z)}^{\rm II}(d,V)$ is clearly found in Fig.~\ref{mv}, 
i.e., $\delta M_{y(z)}^{\rm II}(d,V)$ showing non-zero values only at $V=m\hbar\omega_{\rm S}/2e$.

\section{Discussion}\label{sec:discussion}

Finally, we shall approximately estimate 
the amplitude of the magnetization induced inside the $N$. 
As shown in Figs.~\ref{m1}--\ref{m3}, the magnetization in the $N$ has a finite value in the length scale of $\xi_{\rm D}$. 
In dirty normal metals, $\xi_{\rm D}$ is in a range of several dozen to several hundred 
nanometers since $D_{\rm N}$ is about 0.01--0.1 $\rm m^{2}$/s 
and $T_{\rm C}$ is assumed to be a few kelvins~\cite{deutscher}. 
As indicated in Figs.~\ref{m1}--\ref{mq}, and \ref{mv}, the amplitude of the 
magnetization is estimated to be one to two orders smaller than 
$M_0=g \mu_{\rm B} N_{\rm F} k_{\rm B} T_{\rm C} \xi_{\rm D}^{2}/\xi_{\rm F2}^{2}$. 
When we use a typical set of parameters, i.e., the density of states at the Fermi energy 
\begin{equation}
N_{\rm F}=\frac{1}{4\pi^2}\left( \frac{2m}{\hbar^2} \right)^{3/2}\varepsilon_{\rm F}^{1/2} 
\approx 5.4 \times10^{27}\,\ {\rm eV^{-1}} {\rm m^{-3}} 
\end{equation} 
with the Fermi energy 
$\varepsilon_{\rm F}\approx5$ eV~\cite{book-ssp} ($m$: the electron mass), 
$T_{\rm C}=9$ K for Nb~\cite{deutscher}, $\xi_{\rm D}=100$ nm,  and $\xi_{\rm F2}=5$ nm~\cite{deutscher, kontos-prl}, we can estimate that 
$M_{0}$ is approximately 31000 $\rm{A/m}$. 
It is therefore expected that the magnetization induced inside the $N$ can be 
detected with the magnetization measurement by SQUID~\cite{book-mm}.

\section{Summary}\label{sec:summary}

We have calculated the magnetization inside the $N$ in the $S$/$F$1/$N$/$F$2/$S$ 
Josephson junction based on the the quasiclassical Green's function method in the diffusive 
transport limit. 
By solving the Usadel equation, 
we have found that finite magnetization is induced inside the $N$. 
We have shown that the magnetization is due to the odd-frequency STCs formed by 
electrons of equal and opposite spins, which are induced by the proximity effect in the 
{\it S}/{\it F}1/{\it N}/{\it F}2/{\it S} junction. 
Fixing the magnetization in $F$2 along the $z$ direction perpendicular to the junction direction 
($x$ direction), we have shown that i) the $x$ component of the magnetization in the $N$ is 
always zero, ii) the $y$ component is exactly zero when the magnetization direction between 
$F$1 and $F$2 is collinear, and iii) the $z$ component is generally finite for any magnetization 
direction between $F$1 and $F$2. 

Decomposing the induced magnetization into $\theta$ independent and dependent parts, we have 
found that the $\theta$ independent part of the magnetization decays slowly with increasing 
the thickness of the $N$, whereas the $\theta$ dependent part of the magnetization decays rather rapidly. 
While the $\theta$ independent part of the magnetization is generally induced 
even in the $S$/$F$ junctions due to the proximity effect, the $\theta$ dependent part of the 
magnetization results from the finite coupling between the two $S$s in the $S$/$F$1/$N$/$F$2/$S$ 
Josephson junction.  
We have also found that the time averaged magnetization in the $N$ exhibits discontinuous peaks 
at particular values of DC voltage when DC and AC voltages are both applied to the 
$S$/$F$1/$N$/$F$2/$S$ junction, 
implying that the AC magnetization oscillation can be converted into the DC component. 
We have discussed that the magnetization induced inside the $N$ can be large enough 
to be observed in typical experimental settings. 
It is therefore expected that a Josephson junction composed of ferromagnetic metallic multilayers 
such as the one studied here can have a promising potential for low Joule heating spintronics 
devices, where the magnetization can be controlled by varying the superconducting phase 
difference $\theta$.

\section*{ACKNOWLEDGMENTS} 
This work is supported by Grant-in-Aid for Research Activity Start-up (No. 25887053) from the 
Japan Society for the Promotion of Science and also in part by RIKEN iTHES Project.

\appendix{

\section{Spatial dependence of anomalous Green's functions inside normal metal}\label{app:agf}

In this Appendix, we shall discuss the spatial dependence of the anomalous Green's functions inside 
the $N$. The analytical solutions are obtained by solving the linearized Usadel equation 
(see Sec.~\ref{sec:agf}) and are given in Eqs.~(\ref{fns})--(\ref{fntz}). 
Figure~\ref{f-x} shows the typical results of the anomalous Green's junctions inside 
the {\it N} for three different magnetization alignment between $F$1 and $F$2, parametrized 
by $\varphi$ (see Fig.~\ref{sfnfs-gm}).
As shown in Fig.~\ref{f-x}(a), $f_{s}^{\rm N}(x)$ does not depend on $\varphi$ and 
exhibits symmetric behavior with respect to $x$ about the center of the $N$. 
The $\varphi$ independence 
is simply because $f_{s}^{\rm N}(x)$ represents the SSC which can be induced even without $F$ 
layers [see Eq.~(\ref{fns})]. 

Figure~\ref{f-x}(b) shows the spatial dependence of the anomalous Green's function 
$f_{ty}^{\rm N}(x)$, corresponding to the STC with $|S_{z}|=1$. For the collinear magnetization 
alignment, i.e., $\varphi =0$ or $\pi$, $f_{ty}^{\rm N}(x)$ is exactly zero because $f_{ty}^{\rm N}(x)$ 
is proportional to the $y$ component of the magnetization in $F$1 [see Eq.~(\ref{fnty})]. 
Therefore, the local magnetization density $m_y(x,\theta)$ and thus 
the magnetization $M_y(d,\theta)$ in the $N$ is exactly zero in this case 
(see Figs.~\ref{m1} and \ref{m2}). 
In contrast, $f_{tz}^{\rm N}(x)$, corresponding to the STC with $|S_{z}|=0$, is generally finite, as 
shown in Fig.~\ref{f-x}(c). 
Therefore, the magnetization $M_z(d,\theta)$ inside the $N$ is generally finite, as shown in Figs.~\ref{m1}--\ref{m3}.

It should be noted here that although the analytical solutions in Eqs.~(\ref{fns})--(\ref{fntz}) 
indicate their exponential dependence with respect to $x$, Figs.~\ref{f-x}(b) and \ref{f-x}(c) suggest 
that $f_{ty}^{\rm N}(x)$ for $\varphi\ne0,\pi$ and $f_{tz}^{\rm N}(x)$ vary almost linearly. This seemingly 
linear dependence is simply because of the parameter set chosen in Fig.~\ref{f-x}, where 
$k_{\rm N}\xi_{\rm D}=\sqrt{T/T_{\rm C}} \approx 0.5$ and hence 
$k_{\rm N}(L-x)$ and $k_{\rm N}(x-d_{\rm F1})$ in the exponents are 
no larger than 0.6.

\begin{figure}[t!]
\begin{center}
\vspace{10mm} 
\includegraphics[width=7cm]{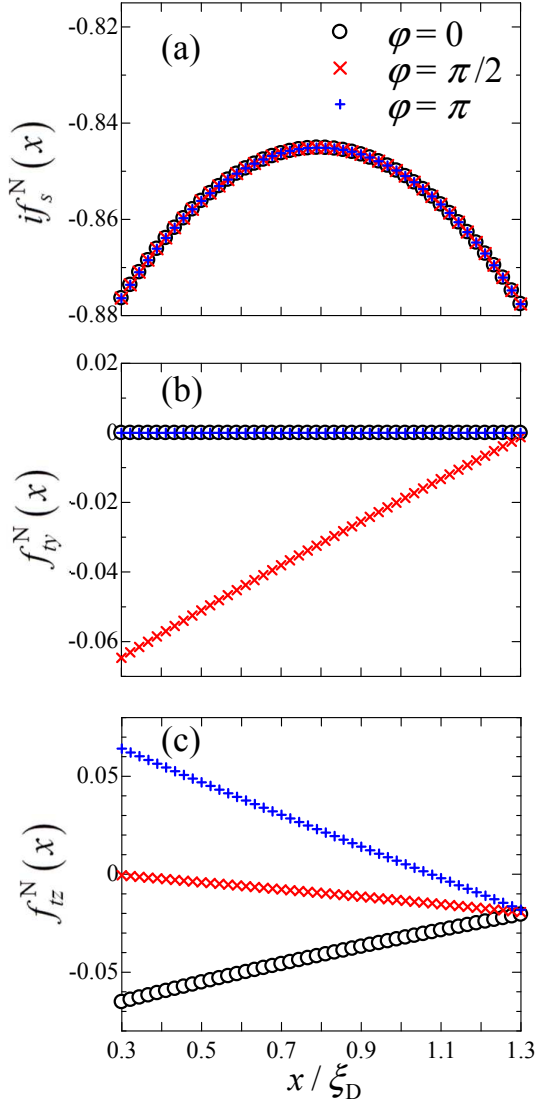}
\caption{ (Color online) Spatial dependence of the anomalous Green's functions, 
(a) $f_{s}^{\rm N}(x)$, (b) $f_{ty}^{\rm N}(x)$, and (c) $f_{tz}^{\rm N}(x)$, inside the 
$N$ at $ \omega_{n}=\pi k_{\rm B} T/\hbar$ 
for $\varphi =0$, $\pi/2$, and $\pi$ indicated in (a). 
We set the thickness $d$ of the $N$ to be $\xi_{\rm D}$ and $\theta_{\rm L}=\theta_{\rm R}=0$. 
Other parameters are the same as in 
Fig.~\ref{m1}. Note that the $N$ layer is located in $0.3\le x/\xi_{\rm D}\le 1.3$ for this parameter set. 
}
\label{f-x}
\end{center}
\end{figure}

\section{Local magnetization density inside normal metal}\label{app:lmd}

In this Appendix, we will first provide the analytical form of the local magnetization density induced 
inside the $N$ and examine the $\varphi$ dependence. 
Within the quasiclassical Green's function method, the local magnetization density $\vec m(x,\theta)$ 
inside the {\it N} is obtained by substituting Eqs.~(\ref{fns})--(\ref{fntz}) into Eq.~(\ref{m}). 
The $y$ component $m_{y}(x,\theta)$ of the local magnetization density can be decomposed into 
$\theta$ independent and dependent parts    
\begin{equation}
	m_{y}(x,\theta) = m_{y}^{\rm I}(x) + m_{y}^{\rm II}(x,\theta)
\label{lmy}, 
\end{equation}
where 
\begin{equation}
	m_{y}^{\rm I}(x) =
		g\mu_{\rm B} \pi k_{\rm B} T N_{\rm F} 
		\frac{h_{\rm ex}^{y}d_{\rm F1}^{2}}{\hbar D_{\rm F}}
		\sum_{\omega_{n}}
			\frac{\Delta^{2}}{E_{\omega_{n}}^{2}}
			K_{\omega_{n}}^{2}(d)
			F_{\omega_{n}}^{2}(x)
\label{lmy1-x} 
\end{equation}
and
\begin{eqnarray}
	m_{y}^{\rm II}(x,\theta) &=&
		-g\mu_{\rm B} \pi k_{\rm B} T N_{\rm F} 
		\frac{h_{\rm ex}^{y}d_{\rm F1}^{2}}{\hbar D_{\rm F}} \nonumber \\
		&\times&
		\sum_{\omega_{n}}
			\frac{\Delta^{2}}{E_{\omega_{n}}^{2}}
			K_{\omega_{n}}^{2}(d)
			F_{\omega_{n}} (x) R_{\omega_{n}} (x) \cos \theta. 
\label{lmy2-x} 
\end{eqnarray}
Here, we have introduced 
%
\begin{equation}
	F_{\omega_{n}} (x) = 
		\sinh[k_{\rm N} (x-L)]
		-\frac{k_{\rm N} d_{\rm F2} }{ \gamma_{\rm F} }
		\cosh[k_{\rm N} (x-L)], 
\end{equation}
and 
\begin{equation}
	R_{\omega_{n}} (x) = 
		\sinh[k_{\rm N} (x-d_{\rm F1})]
		+\frac{k_{\rm N} d_{\rm F1} }{ \gamma_{\rm F} }
		\cosh[k_{\rm N} (x-d_{\rm F1})]. 
\end{equation}
Similarly, the $z$ component $m_{z}(x,\theta)$ of the local magnetization density can be 
decomposed into two parts 
\begin{equation}
	m_{z}(x,\theta) = m_{z}^{\rm I}(x) + m_{z}^{\rm II}(x,\theta)
\label{lmz}, 
\end{equation}
where 
\begin{widetext}
\begin{equation}
	m_{z}^{\rm I}(x) = 
	-g\mu_{\rm B} \pi k_{\rm B} T N_{\rm F} 
	\frac{ h_{\rm ex}^{z} d_{\rm F1}^{2} }{ \hbar D_{\rm F} }
	\sum_{\omega_{n}}
		\frac{ \Delta^{2} }{ E_{\omega_{n}}^{2} }
		K_{\omega_{n}}^{2}(d) F_{\omega_{n}}^{2}(x)
	-g\mu_{\rm B} \pi k_{\rm B} T N_{\rm F} 
	\frac{ h_{\rm ex2} d_{\rm F2}^{2} }{ \hbar D_{\rm F} }
	\sum_{\omega_{n}}
		\frac{ \Delta^{2} }{ E_{\omega_{n}}^{2} }
		K_{\omega_{n}}^{2}(d) R_{\omega_{n}}^{2}(x)
\label{lmz1-x} 
\end{equation}
\end{widetext}
and
\begin{widetext}
\begin{equation}
	m_{z}^{\rm II}(x,\theta) = 
	g\mu_{\rm B} \pi k_{\rm B} T N_{\rm F} 
	\left(
	\frac{ h_{\rm ex}^{z} d_{\rm F1}^{2} }{ \hbar D_{\rm F} }
	+
	\frac{h_{\rm ex2} d_{\rm F2}^{2} }{ \hbar D_{\rm F} }
	\right)
	\sum_{\omega_{n}}
		\frac{ \Delta^{2} }{ E_{\omega_{n}}^{2} }
		K_{\omega_{n}}^{2}(d) F_{\omega_{n}}(x) R_{\omega_{n}}(x) \cos \theta
\label{lmz2-x}. 
\end{equation}
\end{widetext}

\begin{figure}[t!]
\begin{center}
\vspace{10mm} 
\includegraphics[width=6cm]{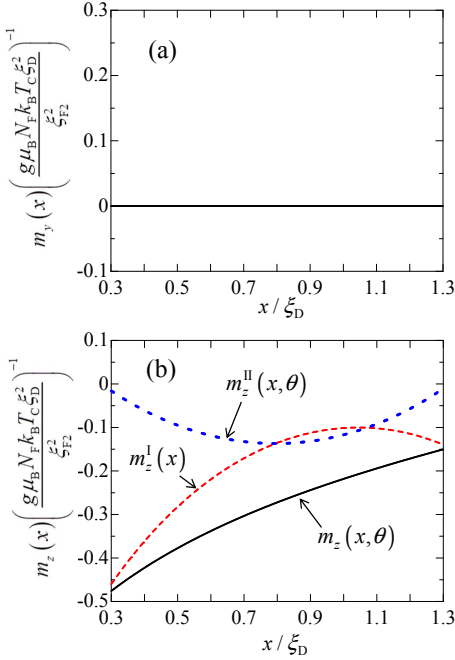}
\caption{ (Color online) (a) The $y$ component $m_{y}(x,\theta)$ and (b) the $z$ component 
$m_{z}(x,\theta)$ of the local magnetization density in the $N$ for $\varphi =0$, 
corresponding to the parallel magnetization configuration between $F$1 and $F$2. 
We set the thickness $d$ of the $N$ to be $\xi_{\rm D}$ and 
other parameters are the same as in Fig.~\ref{m1}. 
For comparison, $m_{z}^{\rm I}(x)$ and $m_{z}^{\rm II}(x,\theta)$ are also plotted separately in (b). 
}
\label{mx1}
\end{center}
\end{figure}

\begin{figure}[t!]
\begin{center}
\vspace{10mm} 
\includegraphics[width=6cm]{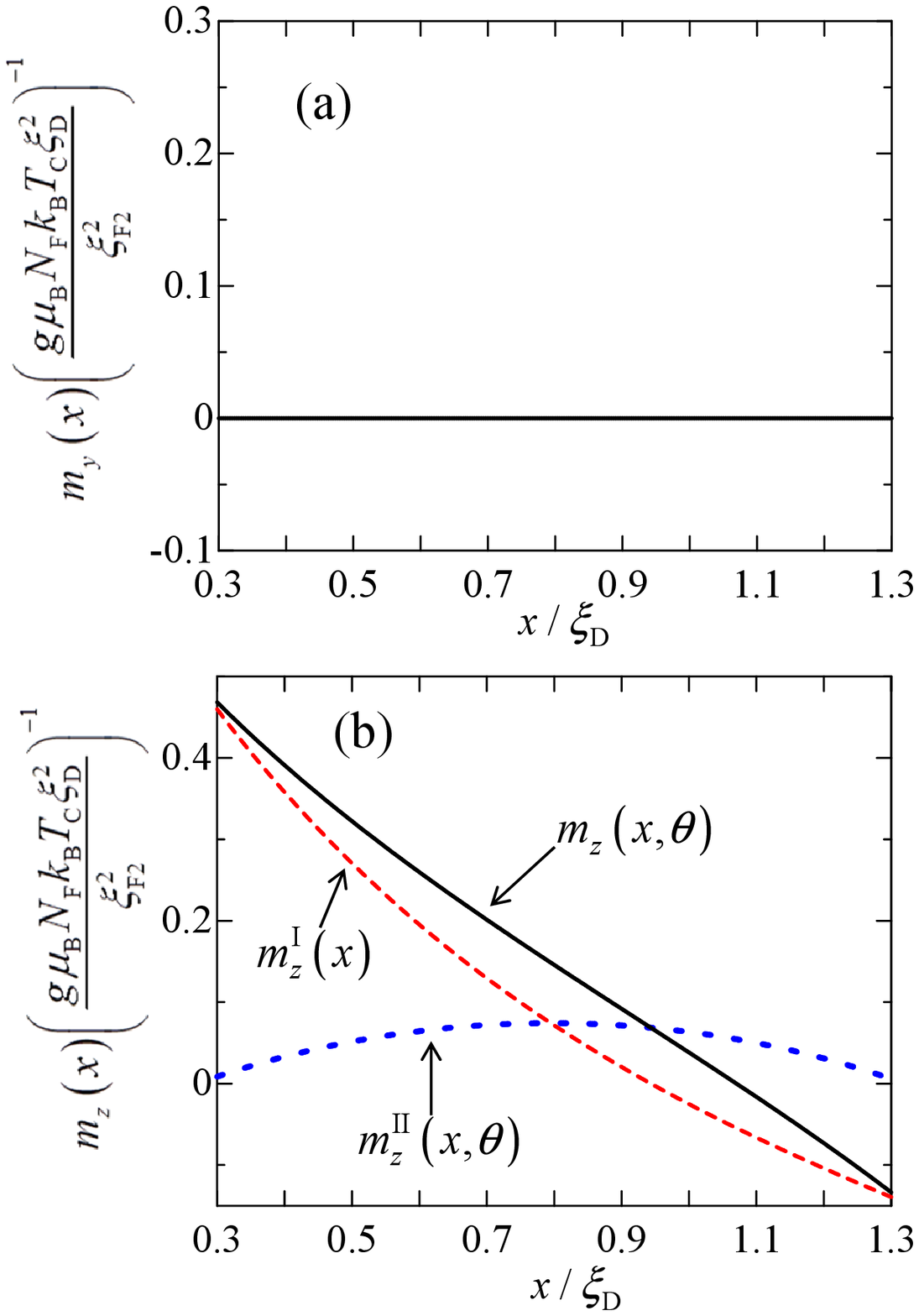}
\caption{ (Color online) (a) The $y$ component $m_{y}(x,\theta)$ and (b) the $z$ component 
$m_{z}(x,\theta)$ of the local magnetization density in the $N$ for $\varphi =\pi$, 
corresponding to the antiparallel magnetization configuration 
between $F$1 and $F$2. 
Other parameters are the same as in Fig.~\ref{mx1}. 
For comparison, $m_{z}^{\rm I}(x)$ and $m_{z}^{\rm II}(x,\theta)$ are also plotted separately in (b). 
}
\label{mx3}
\end{center}
\end{figure}

\begin{figure}[t!]
\begin{center}
\vspace{10mm} 
\includegraphics[width=6.5cm]{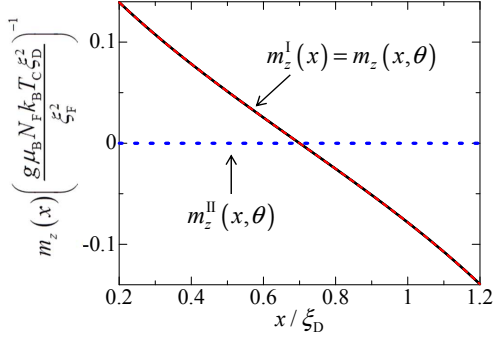}
\caption{ (Color online) The $z$ component 
$m_{z}(x,\theta)$ of the local magnetization density in the $N$ for $\varphi =\pi$, 
corresponding to the antiparallel magnetization configuration 
between $F$1 and $F$2. 
We set $d_{\rm F1}/\xi_{\rm D} = d_{\rm F2}/\xi_{\rm D} = 0.2$ and 
$h_{\rm ex1}/\Delta_{0}=h_{\rm ex2}/\Delta_{0}=20$. 
Other parameters are the same as in Fig.~\ref{mx1}. 
For comparison, $m_{z}^{\rm I}(x)$ and $m_{z}^{\rm II}(x,\theta)$ are also plotted separately. 
}
\label{mx4}
\end{center}
\end{figure}

\begin{figure}[t!]
\begin{center}
\vspace{10mm} 
\includegraphics[width=6cm]{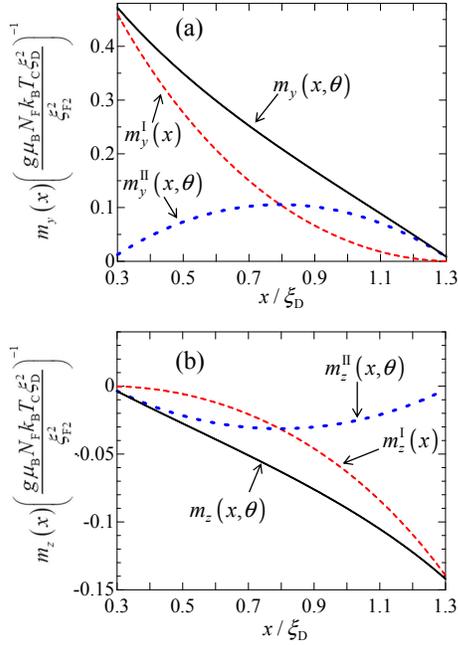}
\caption{ (Color online) (a) The $y$ component $m_{y}(x,\theta)$ and (b) the $z$ component 
$m_{z}(x,\theta)$ of the local magnetization density in the $N$ for $\varphi =\pi/2$, 
corresponding to the case where the magnetization in $F$1 is perpendicular to that in $F$2. 
Other parameters are the same as in Fig.~\ref{mx1}. 
For comparison, $m_{y(z)}^{\rm I}(x)$ and $m_{y(z)}^{\rm II}(x,\theta)$ are also plotted separately. 
}
\label{mx2}
\end{center}
\end{figure}

Figure~\ref{mx1} shows the numerical results for $\varphi=0$ where the magnetizations 
between {\it F}1 and {\it F}2 are parallel. 
As shown in Fig.~\ref{mx1}(a), the $y$ component $m_{y}(x,\theta)$ of the local magnetization 
density is exactly zero because $f_{ty}^{\rm N}(x)$ contributing to $m_{y}(x,\theta)$ is zero in 
the parallel magnetization configuration. 
On the other hand, the $z$ component $m_{z}(x,\theta)$ of the local magnetization density 
has a finite value, as shown in Fig.~\ref{mx1}(b), since $f_{tz}^{\rm N}(x)$ contributing to 
$m_{z}(x,\theta)$ is nonzero in the parallel magnetization configuration. 
Furthermore, the induced local magnetization density $m_{z}(x,\theta)$ is found to be negative, 
i.e., pointing the opposite direction to the magnetizations in $F$1 and $F$2. 
It should also be noticed that both $m_{z}^{\rm I}(x)$ and $m_{z}^{\rm II}(x,\theta)$ exhibit generally 
nonmonotonic behavior with respect to $x$.

Figure~\ref{mx3} shows the numerical results for $\varphi=\pi$ where the magnetizations 
between {\it F}1 and {\it F}2 are antiparallel. 
As shown in Fig.~\ref{mx3}(a), the $y$ component $m_{y}(x,\theta)$ of the local magnetization 
density is exactly zero since $f_{ty}^{\rm N}(x)$ contributing to $m_{y}(x,\theta)$ is zero 
also in the antiparallel magnetization configuration. 
On the other hand, as shown in Fig.~\ref{mx3}(b), the $z$ component $m_{z}(x,\theta)$ of the 
local magnetization density has a finite value since $f_{tz}^{\rm N}(x)$ contributing to 
$m_{z}(x,\theta)$ is nonzero in the antiparallel 
magnetization configuration. 
Furthermore, as opposed to the case for $\varphi=0$, the induced local magnetization 
density $m_{z}(x,\theta)$ changes the sign from positive to negative with increasing $x$. 
Note also that $m_{z}^{\rm II}(x,\theta)$ is exactly zero for the special case when 
$d_{\rm F1}$ =$d_{\rm F2}$ and $|h_{\rm ex}^{z}|$ = $|h_{\rm ex2}|$, as shown in Fig.~\ref{mx4}, 
and thus the local magnetization density is no longer dependent on $\theta$.

Finally, Fig.~\ref{mx2} shows the results for $\varphi=\pi/2$ where the magnetization in $F$1 is 
perpendicular to that in $F$2. 
As shown in Fig.~\ref{mx2}(a), the $y$ component $m_{y}(x,\theta)$ of the local magnetization 
density is now finite because 
$f_{ty}^{\rm N}(x)$ contributing to $m_{y}(x,\theta)$ is nonzero in this case [see Fig.~\ref{f-x}(b)]. 
Similarly to the previous cases for $\varphi=0$ and $\pi$, the $z$ component $m_{z}(x,\theta)$ of 
the local magnetization density is also finite [Fig.~\ref{mx2}(b)]. 

}

{}
\end{document}